\documentclass[12pt]{article}
\usepackage[top=30truemm, bottom=30truemm, left=25truemm, right=25truemm]{geometry}
\usepackage{amsmath,amssymb,ascmac,braket,bm,mathrsfs,amsthm,amsfonts,hyperref}
\usepackage[dvipdfmx]{graphicx}
\usepackage{cite}
\usepackage{bmpsize}
\usepackage{here}
\usepackage{slashed}
\usepackage{ulem}

\newcommand{\del}{\partial}
\newcommand{\nn}{\nonumber}
\newcommand{\ol}{\overline}
\usepackage{color}

\begin{document}

\begin{titlepage}

\begin{flushright}
{\tt 
IPMU18-0149
}
\end{flushright}

\vskip 1.35cm
\begin{center}

{\large
{\bf
  Loop corrections to dark matter direct detection \\
  in a pseudoscalar mediator dark matter model
}}

\vskip 1.2cm

Tomohiro Abe$^{1,2}$,
Motoko Fujiwara$^3$, and 
Junji Hisano$^{2,3,4}$

\vskip 0.4cm

{\it $^1$ Institute for Advanced Research, Nagoya University, \\
  Furo-cho Chikusa-ku, Nagoya, Aichi, 464-8602 Japan} \\[3pt]
{\it $^2$ Kobayashi-Maskawa Institute for the Origin of Particles and the Universe,\\
 Nagoya University, Furo-cho Chikusa-ku, Nagoya, Aichi, 464-8602 Japan}\\ [3pt]
{\it $^3$ Department of Physics, Nagoya University, \\
             Furo-cho Chikusa-ku, Nagoya, Aichi, 464-8602 Japan}\\[3pt]
{\it $^4$ Kavli IPMU (WPI), UTIAS, University of Tokyo, Kashiwa, Chiba 277-8584, Japan}

\date{\today}

\vskip 1.5cm

\begin{abstract}
If dark matter (DM) is a fermion and its interactions with the standard model particles are mediated by pseudoscalar particles, the tree-level amplitude for the DM-nucleon elastic scattering is suppressed by the momentum transfer in the non-relativistic limit. At the loop level, on the other hand, the spin-independent contribution to the cross section appears without such suppression. Thus, the loop corrections are essential to discuss the sensitivities of the direct detection experiments for the model prediction. 
The one-loop corrections were investigated in the previous works. However, the two-loop diagrams give the leading order contribution to the DM-gluon effective operator ($\bar{\chi} \chi G^{a}_{\mu \nu} G^{a \mu \nu} $) and have not been correctly evaluated yet. Moreover, some interaction terms which affect the scattering cross section were overlooked. In this paper, we show the cross section obtained by the improved analysis and discuss the region where the cross section becomes large. 
\end{abstract}

\end{center}
\end{titlepage}
\tableofcontents
\clearpage

\clearpage
\section{Introduction}
Weakly Interacting Massive Particles (WIMPs) are popular dark matter (DM) candidates and
often appear in models beyond the standard model (BSM).
A variety of models have been studied in the literature. 
They typically predict scattering processes between DM and nucleon
with a sizable cross section which can be detected experimentally.
There are many DM direct detection experiments such as the LUX~\cite{1705.03380}, PandaX-II~\cite{Cui:2017nnn}, and XENON1T experiments~\cite{1805.12562}.
The significant DM signals have not been reported yet, 
and these experiments give severe upper bounds on the DM-nucleon spin-independent (SI) scattering cross section ($\sigma_{\text{SI}}$).
This fact gives a strong constraint for the parameter space of the models which predict a WIMP as a DM candidate.

A pseudoscalar coupling with fermion DM is a simple way to avoid these strong constraints from the DM direct detection experiments~\cite{
1609.09079, 1612.06462},
\begin{align}
 \bar{\chi} i \gamma^5 \chi s, \label{eq:chi-gamma5-chi}
\end{align}
where $\chi$ is a fermion as a DM candidate and $s$ is a scalar mediator connecting the DM and the standard model (SM) sector.
In the non-relativistic limit, this interaction term predicts the suppression of the tree-level DM-nucleon scattering amplitude by the momentum transfer.
On the other hand, the amplitude for DM annihilation processes is predicted as $s$-wave.
Therefore the interaction term in Eq.~\eqref{eq:chi-gamma5-chi} has desirable features for WIMP models, 
namely models can evade the strong constraints from the DM direct detection experiments 
while keeping the annihilation cross section to explain the amount of the DM in our universe as a thermal relic abundance.

A pseudoscalar mediator model~\cite{1404.3716} is one of the simplest models which predict the interaction term in Eq.~\eqref{eq:chi-gamma5-chi},
 and its phenomenology has been widely studied~\cite{1509.01110, 1611.04593, 1701.07427, 1705.09670, 1711.02110, 1712.03874, 1803.01574, 1804.02120}.
In the model, $\chi$ is a gauge singlet fermion, and $s$ is a gauge singlet pseudoscalar.
The Higgs sector is extended into the two-Higgs doublet models (THDMs) 
to make the gauge singlet pseudoscalar interact with the SM sector at the renormalizable level.
CP invariance is assumed in the DM and the mediator sectors.\footnote{
If the DM and/or the mediator sector breaks the CP invariance,
the mediator can have renormalizable interactions with the SM sector 
without extending the Higgs sector into the THDMs, see for example~\cite{1408.4929, 1411.1335, 1701.04131, 1702.07236}.
}

In this model, $\sigma_{\text{SI}}$ is generated at the loop level~\cite{1404.3716, 1711.02110, 1803.01574, 1804.02120}.
Although $\sigma_\text{SI}$ is suppressed by the loop factors 
and is smaller than the current upper bounds from the direct detection experiments,
it can be larger than the neutrino floor~\cite{1307.5458} and can be detected by the future DM direct detection experiments.
Therefore it is essential to evaluate the cross section at the loop level.
However, 
$\sigma_{\text{SI}}$ has been calculated without including some relevant interaction terms in~\cite{1404.3716, 1711.02110, 1803.01574, 1804.02120}.
Moreover, two-loop diagrams which induce the DM-gluon effective operator, $\bar{\chi} \chi G_{\mu \nu}^a G^{a  \mu \nu}$, 
have not been correctly calculated as was mentioned in~\cite{1711.02110}.

In this paper, we give a complete set for the leading order calculations in the pseudoscalar mediator DM model.\footnote{
$\chi$ is  a Dirac fermion in~\cite{1404.3716, 1711.02110, 1803.01574, 1804.02120}. 
In this paper, however, we consider $\chi$ as a Majorana 
fermion to make the model much simpler. 
Qualitative features are the same both in the Dirac and the Majorana cases.
}
We take into account all of the renormalizable interaction terms.
We find that quartic interaction terms between the pseudoscalar and the SM Higgs bosons, which have been ignored in~\cite{1404.3716, 1711.02110}, 
are important to enhance $\sigma_{\text{SI}}$. 
As a result of the enhancement, 
the model can be detected by the XENONnT~\cite{1512.07501}, LZ~\cite{1611.05525}, and DARWIN experiments~\cite{1606.07001}
in some parameter regions.
We also calculate the relevant two-loop diagrams for the DM-gluon effective operators.
In~\cite{1711.02110}, the contribution was estimated from the one-loop box diagrams 
by using a relation between a heavy quark scalar-type operator and a gluon scalar-type operator~\cite{Shifman:1978zn} 
without justification.
We find that the contributions from the charm and bottom quarks were underestimated, 
while the contribution from the top quark was overestimated in~\cite{1711.02110}. 
In the end, we clarify that the contribution from the box diagrams is subdominant and the triangle diagrams dominate the scattering process.

The structure of this paper is as follows:
In Sec.~\ref{sec:model}, we introduce the gauge invariant renormalizable model which contains the pseudoscalar mediators~\cite{1404.3716}. 
In Sec.~\ref{sec:eo}, we derive the effective operators which induce the DM-nucleon SI scattering. 
In Sec.~\ref{sec:result}, we show our numerical results. 
We compare our result with the previous one in~\cite{1711.02110}, and then search the parameter space 
where $\sigma_{\text{SI}}$ becomes large enough to reach the prospects of the direct detection experiments. 
Our conclusions are in Sec.~\ref{sec:conclusion}. 
In Appendix \ref{sec:3scalar}, we show explicit formulas for scalar trilinear couplings which are defined in Sec.~\ref{sec:model}. 
In Appendix \ref{sec:boxdetail}, we write the details of the derivation of the effective operators for SI scattering between DM and quarks/gluon.  
In Appendix \ref{sec:appendix_loop}, we define the loop functions used in Sec.~\ref{sec:eo}.

\section{Model}
\label{sec:model}
In this section, we briefly review the pseudoscalar mediator DM model~\cite{1404.3716}. 
The model contains a gauge singlet Majorana fermion $\chi$ as a DM candidate
and a gauge singlet pseudoscalar boson $a_0$ as a mediator field.
The DM can be expressed using Weyl spinor $\psi$ as follows: 
\begin{align}
    \chi
    =
    \begin{pmatrix}
        \psi  
        \\
        \psi^\dagger
    \end{pmatrix}.
\end{align}
The Higgs sector is also extended into a THDM, 
which contains two SU(2)$_L$ doublet scalar fields $H_j$ $(j  =  1, 2)$ with a hypercharge $Y  =  1/2$.

We assume a $Z_2$ symmetry to stabilize the DM candidate. 
Under this $Z_2$ symmetry, $\chi$ is odd, 
and all the other fields are even.
The interaction terms of the DM and scalar fields are given by
\begin{align}
 {\cal L} 
\supset&  
+ i \frac{ g_\chi }{ 2 }  a_0  \bar{ \chi }  \gamma^5  \chi 
- \left(V_{\rm THDM}  + V_{a_0} + V_{\rm port} \right),  
\label{eq:Lag}
\end{align}
where
\begin{align}
 V_{\rm THDM}  
=&  m_{1}^2  H_1^\dagger H_1  +  m_{ 2 }^2  H_2^\dagger  H_2   
    -m_{ 3 }^2  \left( H_1^\dagger H_2 + { \rm h.c. } \right)  \nn   \\
 &   +  \frac{\lambda_1}{2}  (H_1^\dagger H_1)^2  
 +  \frac{\lambda_2}{2}  (H_2^\dagger H_2)^2  
 +  \lambda_3  ( H_1^\dagger H_1 )  ( H_2^\dagger H_2 )  
 +  \lambda_4  ( H_1^\dagger  H_2 )  ( H_2^\dagger H_1 ) 
 \nn   \\
 &   +  \frac{ \lambda_5 }{ 2 }  \left[ ( H_1^\dagger  H_2 )^2  + { \rm h.c. }  \right], 
 \label{eq:poteTHDM}
\\
 V_{a_0} =& \frac{1}{2}  m^2_{a_0}  a_0^2  +  \frac{ \lambda_{a_0} }{ 4 }  a_0^4 , \label{eq:va0}\\
 V_{\rm port} =&  \kappa(i a_0 H_1^\dagger H_2 + {\rm h.c.} )  +  c_1  a_0^2  H_1^\dagger  H_1  +  c_2  a_0^2  H_2^\dagger  H_2.
    \label{eq:darkint}
\end{align}
Here we assume CP invariance in Eq.~\eqref{eq:Lag}, and therefore all the parameters in Eq.~\eqref{eq:Lag} are real.
We also assume a softly broken $Z_4$ symmetry to avoid flavor changing Higgs couplings at the tree-level.
This symmetry is an extension of the softly broken $Z_2$ symmetry often assumed in studies of the THDMs~\cite{PhysRev.D41.3421,
hep-ph/9401311, 0902.4665} to avoid flavor changing scalar couplings~\cite{PhysRev.D15.1958}.\footnote{
For the analysis without any discrete symmetry to forbid the flavor changing scalar couplings, see~\cite{1508.05716}.
}
Under this $Z_4$ symmetry, each field is transformed as follows:
\begin{align}
    \psi                &\mapsto                i  \psi,    \\
    a_0                &\mapsto                -  a_0,   \\
    H_1               &\mapsto                H_1,      \\
    H_2               &\mapsto                -  H_2.
\end{align}  
For the SM fermions, there are four variations in charge assignments as summarized in Table~\ref{tab:SMZ4}. 
    \begin{table}[]
        \centering
        \begin{tabular}{|  l  || c  | c | c | c |}\hline
         &  $Q_L^i$, $L_L^i$  &  $u_R^i$  &  $d_R^i$  &  $e_R^i$  
         \\\hline \hline
        Type-I
        &  $+$  &  $-$  &  $-$  &  $-$  
        \\\hline
        Type-II
        &  $+$  &  $-$  &  $+$  &  $+$  
        \\\hline
        Type-X
        &  $+$  &  $-$  &  $-$  &  $+$  
        \\\hline
        Type-Y
        &  $+$  &  $-$  &  $+$  &  $-$  
        \\\hline
        \end{tabular}
        \caption{The charge assignments of the $Z_4$ symmetry for the SM fermions,  
                      where $Q_L^i$, $L_L^i$, $u_R^i$, $d_R^i$, and $e_R^i$ are 
                      the $i$-th generation of the left-handed quark, 
                      the left-handed lepton, 
                      the right-handed up-type quark, 
                      the right-handed down-type quark, and 
                      the right-handed charged lepton, respectively ($i  =  1, 2, 3$). 
                      } 
        \label{tab:SMZ4}
    \end{table}
This $Z_4$ symmetry is softly broken by the DM mass term as well as the $m_3^2$ term in $V_{\rm  THDM}$.\footnote{
This $Z_4$ symmetry is different from the $Z_2$ symmetry for the DM stability which we mentioned above. }

As can be seen from the scalar potential,
$a_0$ is mixed with the CP-odd scalar in the THDM sector after the electroweak symmetry breaking, and thus
the DM interacts with the SM particles by exchanging the pseudoscalar particles.
This interaction structure is crucial to evade the direct detection constraints as we mentioned in the introduction.

Note that the interaction terms proportional to $c_i$ $(i = 1, 2)$ were not included in the analysis of~\cite{1404.3716} and~\cite{1711.02110}. 
Although the existence of these interaction terms was pointed out in~\cite{1803.01574}, they also neglected these terms in their analysis.
As we will see in later, however, the effect of the $c_2$ term plays an important role in
a DM-nucleon scattering process for the DM direct detection experiments.

The field definitions of the Higgs doublets are as follows:
\begin{align}
H_j&=
  \begin{pmatrix}
    \phi^+_j\\
    \frac{1}{\sqrt{2}}(v_j+\rho_j + i \eta_j)
  \end{pmatrix}
~~~
( j = 1, 2 ),
\end{align}
where $v_j$ is the vacuum expectation value of each doublet field.
We introduce $\tan \beta$ as the ratio of $v_1$ and $v_2$, 
\begin{align}
\tan\beta = 
\frac{v_2}{v_1}, ~~~~~
        v^2 = v_1^2 + v_2^2,
\end{align}
where $v \simeq 246 \ \text{GeV}$, the vacuum expectation value of the SM Higgs boson.
For simplicity, we use $t_\beta$ as an abbreviation for $\tan\beta$. 
We assume that $a_0$ has no vacuum expectation value. 
Then, the scalar mass eigenstates are given from the weak eigenstates as follows: 
\begin{align}
 \begin{pmatrix}
  G^\pm  \\  H^\pm
 \end{pmatrix}
&=
 \begin{pmatrix}
  \cos\beta &\sin\beta   \\
  -\sin\beta& \cos\beta
 \end{pmatrix}
 \begin{pmatrix}
  \phi^\pm_1  \\  \phi^\pm_2
 \end{pmatrix}
, \\
 \begin{pmatrix}
  H  \\  h
 \end{pmatrix}
&=
 \begin{pmatrix}
  \cos \alpha  &  \sin \alpha  \\
  -\sin\alpha& \cos\alpha
 \end{pmatrix}
 \begin{pmatrix}
  \rho_1  \\  \rho_2
 \end{pmatrix}
,\\
 \begin{pmatrix}
  G_0  \\  A \\ a
 \end{pmatrix}
&=
 \begin{pmatrix}
  1 & 0 & 0 \\
  0 & \cos\theta & \sin\theta \\
  0 & -\sin\theta & \cos\theta \\
 \end{pmatrix}
 \begin{pmatrix}
  \cos\beta &\sin\beta & 0 \\
  -\sin\beta& \cos\beta & 0 \\
  0 & 0 & 1
 \end{pmatrix}
 \begin{pmatrix}
  \eta_1  \\  \eta_2 \\ a_0
 \end{pmatrix},
\end{align}
where $G^{\pm}$ and $G_0$ are would-be Nambu Goldstone bosons for $W^{\pm}$ and $Z$, respectively.
There is the following relation between $\theta$ and $\kappa$.\footnote{
Note that the overall sign disagrees with the expression shown in~\cite{1404.3716} and~\cite{1711.02110} when $\sin 2 \theta$ is rearranged to be $\tan 2 \theta$. } 
\begin{align}
  \sin  2  \theta    &=    \frac{ -  2  \kappa  v }{ m_A^2  -  m_a^2 },
\end{align}
where $m_a$ and $m_A$ are the mass eigenvalues of the pseudoscalar states $a$ and $A$, respectively.

In the following discussion, 
we take the alignment limit $\sin (\beta  -  \alpha)  \to  1$ 
where arbitrary values of $t_\beta$ are allowed from the latest LHC constraints~\cite{ATLAS:2018doi, 1809.10733}.\footnote{
A large difference of $\beta-\alpha$ from $\pi/2$ is disfavored.
The measurements of the Higgs couplings give the most stringent bound on $\beta-\alpha$.
For example, 
$|\cos(\beta-\alpha)|  \lesssim  0.01$ is allowed for all the types of the THDM
with $t_\beta=10$~\cite{ATLAS:2018doi}. 
We evaluated $\sigma_{\text{SI}}$ for $|\cos(\beta-\alpha)|  \lesssim  0.01$ 
and found that $\sigma_{\text{SI}}$ takes the maximum value in the alignment limit. 
We also found that $\sigma_{\text{SI}}$ can be 0.4 times smaller than that of the alignment limit
for the type-II case.
Since we are interested in the parameter points where $\sigma_{\text{SI}}$ becomes large,
we took the alignment limit throughout our analysis.
}
Under this limit, 
the interactions of $h$ are similar to those of the SM-Higgs boson.
We also assume $m_H ~= ~m_{H_\pm}~= ~m_A$ to avoid constraints from the electroweak precision measurements by enhancing the custodial symmetry.
Under this setup, the free parameters of this model are as follows: 
\begin{align}
    \{ m_\chi, ~g_\chi,~m_a,~m_A,~\theta,~t_\beta,~c_1,~c_2 \}.
\end{align}

In the following, 
we briefly review the constraints on these parameters. 
In this paper, 
we focus on the parameter regions which are allowed for all the THDM types. 
The lower bound on the charged Higgs boson mass is $m_{H_\pm}  >  580$~GeV from $B  \to  X_s  \gamma$ for the type-II THDM~\cite{1702.04571}, 
and thus we take $m_H  =  m_{H_\pm}  =  m_A  =  600$~GeV as the benchmark point. 
With this choice, we find that $8  \lesssim  t_\beta  \lesssim  15$~\cite{1712.06518, CMS-PAS-HIG-18-005}
and take $t_\beta  =  10$ as the benchmark point. 
As for the light pseudoscalar mass, 
we find $m_a  >  m_h/2$ from the constraint on the Higgs branching ratio~\cite{ATLAS:2018doi, CMS-PAS-HIG-17-031} as discussed in Sec.~\ref{sec:future}. 
Since the loop corrections to $\sigma_{\text{SI}}$ is smaller for the larger $m_a$, 
we focus on $m_a  \lesssim  100$ ~GeV. 
We have checked that $\theta$ is not constrained with these parameters~\cite{1701.07427} 
and simply take the same value as in~\cite{1711.02110}.
We also assume $c_1 = 0$ and $-1  <  c_2  <  1$ in our analysis.
In Table \ref{tab:benchmark_point}, 
we summarize the parameter values and regions considered in this paper. 
As discussed above, 
these values are consistent with the current bounds from the Higgs and flavor observables, 
LHC searches, 
and electroweak precision measurements for all the THDM-types. 
In Sec.~\ref{sec:result}, 
we show the cross section as a function of $m_\chi$ 
by determining $g_\chi$ to realize the thermal relic 
and search the parameter region where the cross section is enhanced. 
{\renewcommand\arraystretch{1.2}
    \begin{table}[]
        \centering
        \begin{tabular}{| c  | c  | c  | c  | c  | c  |}
        \hline
        $m_a$  &  $m_A$  &  $\theta$  &  $t_\beta$  &  $c_1$  &  $c_2$  
        \\
        \hline  
        \hline
        $m_h/2  \sim  100~\text{GeV}$  &  $600$~GeV  &  $0.1$  &  $10$  &  0  &  $-1  \sim  1$
        \\  
        \hline
        \end{tabular}
        \caption{
          The parameter values and regions discussed in this paper. 
          These values are allowed from the bounds of the Higgs and flavor observables, 
          LHC searches, 
          and electroweak precision measurements for all the THDM-types.
          } 
        \label{tab:benchmark_point}
    \end{table}
    }

There are two pseudoscalar mediators, $a$ and $A$. 
The interaction terms between the DM and the mediators are 
\begin{align}
    {\cal L}_{\rm dark} &= i\frac{\xi^\chi_a}{2} a \bar{\chi} \gamma^5 \chi + i \frac{\xi^\chi_A}{2} A \bar{\chi} \gamma^5 \chi  ,
\end{align}
where
\begin{align}
    \xi^{\chi}_a  =  g_{\chi}  \cos  \theta, \quad
    \xi^{\chi}_A  =  g_{\chi}  \sin  \theta.
\end{align}

The scalar trilinear couplings appear from both $V_{ \rm  THDM }$ and $V_{ \rm  port }$ as
\begin{align}
    V_{\rm THDM}  +  V_{\rm port}    \supset &  \sum_{\phi = h, H} \left( -\frac{1}{2}  g_{\phi aa}  \phi  a  a  -  g_{\phi aA}  \phi  a  A  -  \frac{1}{2}  g_{\phi AA}  \phi  A  A  \right).
    \label{eq:trilinear}
\end{align}
These couplings induce the DM-nucleon SI scattering at the loop level
as we will see in Sec.~\ref{sec:eo}.  
The expressions of these couplings are shown in Appendix \ref{sec:3scalar}.

There are four types of the Yukawa structures depending on the $Z_4$ charge assignments of the SM fermions.
\begin{align}
{\cal L}_{\rm Yukawa} 
\supset
\sum_{f}
\left( -\frac{m_f}{v} \right)
\left(
  \xi^f_h h\bar{f}f 
+\xi^f_H H\bar{f}f  
+\xi^f_a a\bar{f}i\gamma^5f
+\xi^f_A A\bar{f}i\gamma^5f
\right)
,
\end{align}
where $f$ indicates the SM fermion and $m_f$ is its mass. 
The rescaling factors $\xi_\phi^f (\phi = h, H, a, A)$ under the alignment limit are shown in Table \ref{tab:yukawaxi}.
As can be seen, the THDM-type dependence appears through the Yukawa couplings of $H, a$, and $A$. 
    \begin{table}[]
        \centering
        \begin{tabular}{|  l  || c c c | c c c | c c c | c c c |}\hline
         & $\xi^u_h$ & $\xi^d_h$ & $\xi^e_h$  
         & $\xi^u_H$ & $\xi^d_H$ & $\xi^e_H$  
         & $\xi^u_a$ & $\xi^d_a$ & $\xi^e_a$
          & $\xi^u_A$ & $\xi^d_A$ & $\xi^e_A$
         \\\hline \hline
        Type-I
        &$  1  $&$  1  $&$  1  $ 
        &$-  \frac{ 1 }{ t_\beta }$ &$-  \frac{ 1 }{ t_\beta }$  &$-  \frac{ 1 }{ t_\beta }$ 
        &$\frac{s_\theta}{t_\beta}$ &$-\frac{s_\theta}{t_\beta}$ &$-\frac{s_\theta}{t_\beta}$
        &$-\frac{c_\theta}{t_\beta}$ &$\frac{c_\theta}{t_\beta}$ &$\frac{c_\theta}{t_\beta}$
        \\\hline
        Type-II
        &$  1  $ &$  1   $&$  1   $
        &$-  \frac{ 1 }{ t_\beta }$ &$t_\beta$ &$t_\beta$
        &$\frac{s_\theta}{t_\beta}$ &$t_\beta s_\theta$ &$t_\beta s_\theta$
        &$-\frac{c_\theta}{t_\beta}$ &$-t_\beta c_\theta$ &$-t_\beta c_\theta$
        \\\hline
        Type-X
        &$  1  $ &$  1  $ &$  1   $
        &$-  \frac{ 1 }{ t_\beta }$ &$-  \frac{ 1 }{ t_\beta }$  &$t_\beta$ 
        &$\frac{s_\theta}{t_\beta}$ &$-\frac{s_\theta}{t_\beta}$ &$t_\beta s_\theta$
        &$-\frac{c_\theta}{t_\beta}$ &$\frac{c_\theta}{t_\beta}$ &$-t_\beta c_\theta$
        \\\hline
        Type-Y
        &$  1  $ &$  1   $&$  1   $
        &$-  \frac{ 1 }{ t_\beta }$ &$t_\beta$ &$-  \frac{ 1 }{ t_\beta }$
        &$\frac{s_\theta}{t_\beta}$ &$t_\beta s_\theta$ &$-\frac{s_\theta}{t_\beta}$
        &$-\frac{c_\theta}{t_\beta}$ &$-t_\beta c_\theta$ &$\frac{c_\theta}{t_\beta}$
        \\\hline
        \end{tabular}
        \caption{The rescaling factors of the Yukawa couplings under the alignment limit,
                      where $s_\theta  =  \sin\theta$ and $c_\theta  =  \cos\theta$. 
                      } 
        \label{tab:yukawaxi}
    \end{table}

\section{Direct Detection}
\label{sec:eo}
If DM is a Majorana fermion,
the relevant effective operators for the evaluation of the DM-nucleon SI scattering cross section are
given by
\begin{align}
 {\cal L}_{\rm  eff } 
=&  
  \frac{1}{2} \sum_{q=u,d,s}  C_q m_q  \bar{\chi}  \chi  \bar{q}  q
 + \frac{1}{2} C_{G}  \left( -  \frac{9  \alpha_s}{8\pi}  \bar{\chi} \chi  G^a_{\mu\nu}  G^{a \mu \nu} \right)
  \nn   \\
  &
  + \frac{1}{2}  \sum_{q=u,d,s,c,b}
   \left[
     C_q^{(1)} \bar{ \chi }  i  \del^\mu  \gamma^\nu  \chi  {\cal O}_{\mu \nu}^q
  +  C_q^{(2)} \bar{ \chi }  i  \del^\mu  i  \del^\nu  \chi {\cal O}_{\mu \nu}^q
 \right],
 \label{eq:eff_ope}
\end{align}
        where ${ \cal  O }_{ \mu  \nu }^q$ is the twist-2 operator for quark $q$,
        \begin{align}
        { \cal  O }_{ \mu  \nu }^q        &=     \frac{i}{2}  \bar{ q }  \left( D_\mu  \gamma_\nu  +  D_\nu  \gamma_\mu  -  \frac{1}{2}  g_{\mu \nu}  \slashed{D} \right)  q.
        \end{align}
We use the following relations to evaluate the SI cross section from these operators~\cite{Hisano:2017jmz},
\begin{align}
 \Braket{N|  m_q  \ol{q}  q  |N}    &=    m_N  f^{N}_{T_q}  ,  \label{eq:mescaq} \\
 - \frac{9 \alpha_s}{8\pi} \Braket{N| G_{\mu \nu}^a  G^{a \mu \nu} |N}    &= m_N f^{N}_{T_G}, \label{eq:mescag}\\
 \Braket{ N | {\cal  O}^q_{\mu  \nu} | N }    &=   
  \frac{1}{m_N}  \left( p_\mu  p_\nu  -  \frac{1}{4}  m_N^2  g_{\mu  \nu} \right) \biggl( q^{N}(2)  +  \ol{q}^{N}(2) \biggr), \label{eq:metwist}
\end{align}
where $N$ stands for a nucleon ($p$, $n$), and $m_N$ is its mass. 
The numerical values of the matrix elements ($f^{N}_{T_q}$)
and the second moments of the parton distribution functions (PDFs) for the quark and anti-quark ($ q^{N}(2)$ and $\ol{q}^{N}(2) $)
are given in Tables~\ref{tab:form} and  \ref{tab:socond_momoent}, respectively.
\begin{table}[]
    \centering
    \begin{tabular}{|c|c|}
    \hline
        \multicolumn{2}{|c|}{For proton}
        \\  \hline
        $f^p_{T_u}$  &  0.0153  
        \\  \hline
        $f^p_{T_d}$  &  0.0191
        \\  \hline
        $f^p_{T_s}$  &  0.0447
        \\
        \hline
    \end{tabular}
    \hspace{2cm}
    \begin{tabular}{|c|c|}
    \hline
        \multicolumn{2}{|c|}{For neutron}
        \\  \hline
        $f^n_{T_u}$  &  0.0110  
        \\  \hline
        $f^n_{T_d}$  &  0.0273
        \\  \hline
        $f^n_{T_s}$  &  0.0447
        \\
        \hline
    \end{tabular}
    \caption{Numerical values of matrix elements which are taken from the default value of \texttt{micrOMEGAs}~\cite{1305.0237}. 
                  The left panel shows the value for the proton, and the right for the neutron. 
                  }
    \label{tab:form}
\end{table}
\begin{table}[]
    \centering
    \begin{tabular}{|c|c||c|c|}
    \hline
        \multicolumn{4}{|c|}{Second moment at $\mu = m_Z$}
        \\  \hline
        $u^p(2)$  &  0.22    &  $\ol{u}^p(2)$  &  0.034
        \\  \hline
        $d^p(2)$  &  0.11    &  $\ol{d}^p(2)$  &  0.036
        \\  \hline
        $s^p(2)$  &  0.026  &  $\ol{s}^p(2)$  &  0.026
        \\  \hline
        $c^p(2)$  &  0.019  &  $\ol{c}^p(2)$  &  0.019
        \\  \hline
        $b^p(2)$  &  0.012  &  $\ol{b}^p(2)$  &  0.012
        \\
        \hline
    \end{tabular}
    \hspace{2cm}
    \caption{Numerical values of the second moments for quark distribution functions for proton. 
                  These values are evaluated at the scale $\mu  =  m_Z$ by using the CTEQ PDFs~\cite{Pumplin:2002vw}. 
                  The values for neutron are given by exchanging up and down quarks in the table. }
    \label{tab:socond_momoent}
\end{table}
The gluon matrix element $(f^{N}_{T_G})$ is given as follows:~\cite{Shifman:1978zn}
\begin{align}
 f^{N}_{T_G} &=  1  -  \sum_{q=u,d,s}  f^{N}_{T_q}.
\end{align}
The values of $q^{N}(2)$ and $\ol{q}^{N}(2)$
are calculated at the scale $\mu  =  m_Z$, where $m_Z$ is $Z$ boson mass.
The DM-nucleon SI scattering cross section is given by
\begin{align}
 \sigma_{\rm  SI}  =  \frac{1}{\pi}  \left( \frac{ m_\chi  m_N }{ m_\chi  +  m_N } \right)^2  |C_N|^2, \label{eq:sigmachin}
\end{align}
where
\begin{align}
 C_N 
=
m_N\left[
\scalebox{0.98}{$\displaystyle 
  \sum_{q=u,d,s} C_q  f_{T_q}^{N}  
 +   C_G  f_{T_G}^{N}  
 +  \frac{ 3 }{ 4 }  \sum_{q=u,d,s,c,b}   \bigl( m_\chi  C_q^{ ( 1 ) }  +  m_\chi^2  C_q^{ ( 2 ) } \bigr)  \bigl( q^{N}(2)  +  \ol{q}^{N}(2) \bigr) 
 $}
\right].
\end{align}
The Wilson coefficients ($C_q$, $C_G$, $C_q^{(1)}$, and $C_q^{(2)}$) are model dependent parts.
In the rest of this section, we calculate these Wilson coefficients at the leading order.

In the pseudoscalar mediator DM model, all of the Wilson coefficients in Eq.~\eqref{eq:eff_ope} are zero at the tree-level.
Diagrams at the one-loop level give the leading order contributions to $C_q$, $C_q^{(1)}$, and $C_q^{(2)}$. 
For $C_G$, the leading order contribution arises at the two-loop level.
Note that the gluon matrix element is defined with the one-loop factor in Eq.~\eqref{eq:mescag},
and thus the contribution from $C_G$ to $\sigma_{\text{SI}}$ is the same order of magnitude as the contributions from the other Wilson coefficients.
For the later convenience, we divide $C_q$ and $C_G$ into the contributions from triangle and box diagrams. 
We introduce the following notations.
\begin{align}
 C_q &= C_q^{\rm tri} + C_q^{ \rm box },\\
 C_G &= C_G^{\rm tri} + C_G^{ \rm box },\\
 C_q^{ ( 1 ) }  &=    C_q^{ ( 1 )  { \rm box } },\\
 C_q^{ ( 2 ) }  &=    C_q^{ ( 2 )  { \rm box } }.
\end{align}

\subsection{Triangle diagrams}
\begin{figure}[]
 \centering
      \includegraphics[width=5cm]{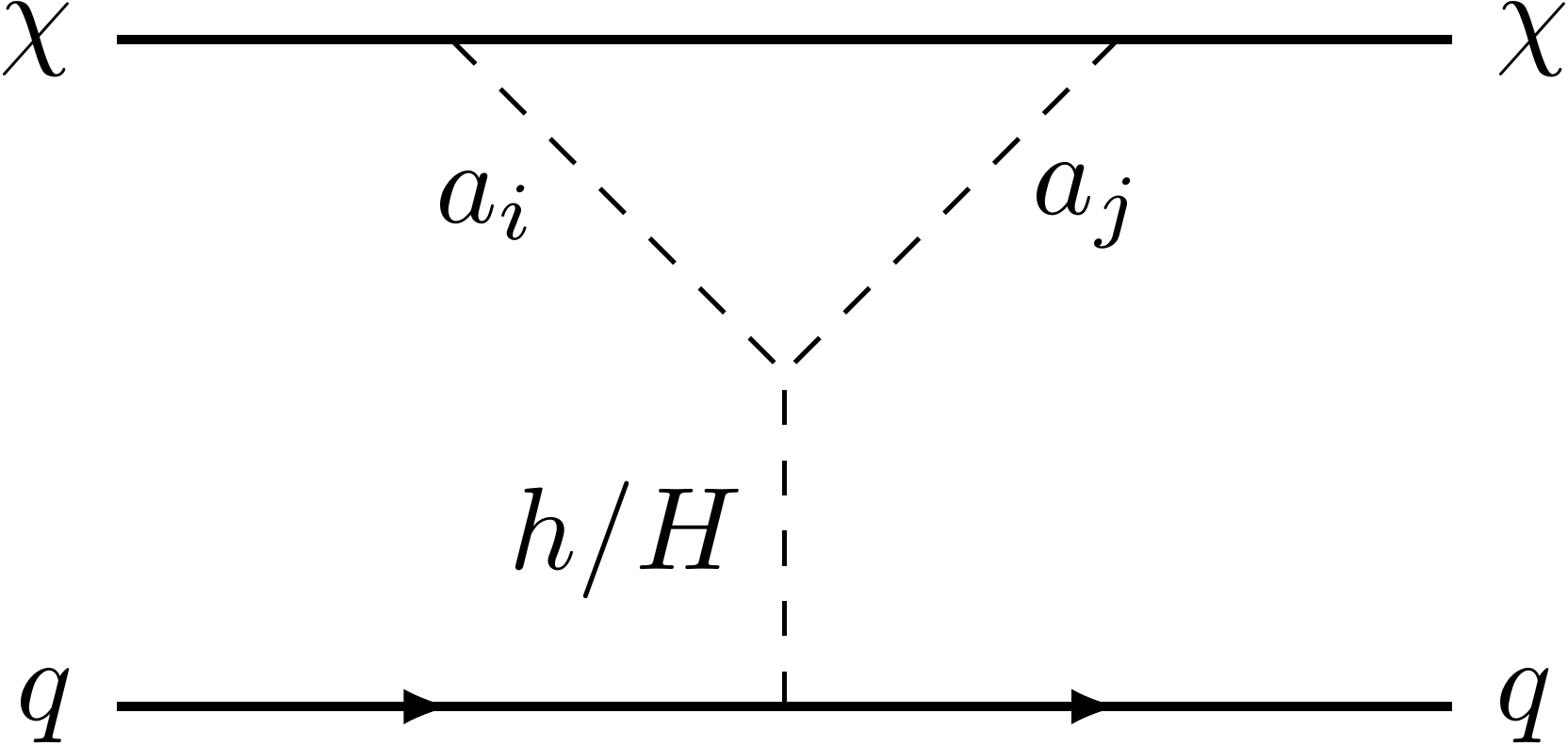}
      \caption{Triangle diagrams which contribute to the DM-nucleon SI scattering. 
              $a_i$ indicates the pseudoscalar mediators, $a$ and $A$.
              $q$ in the external line stands for quarks. 
              The diagrams with each of the light quarks ($u, d, s$) contribute to $\bar{\chi}  \chi  \bar{q}  q$. 
              The diagrams with heavy quarks ($c, b, t$) contribute to $\bar{\chi}  \chi  G_{\mu  \nu}^a  G^{a  \mu  \nu}$. 
              }
    \label{fig:triandbox}
\end{figure}
In the following, we show the effective operators from the triangle diagrams shown in Fig.~\ref{fig:triandbox}.

First, we consider the triangle diagrams with the light quark ($q=u, d, s$) in the external line. 
Each diagram generates the following effective interaction between DM and quark $q$, 
\begin{align}
 {\cal L}_{\rm eff}  &\supset   
 \frac{ 1 }{ 2 }  \sum_{q=u, d, s}  C_{q}^{ \rm tri }  m_q \bar{\chi} \chi \bar{q} q,
\end{align}
where
\begin{align}
 C_{q}^{ \rm tri }  &=   - \sum_{\phi  =  h,  H}  \frac{ \xi^q_\phi }{ m_\phi^2  v }  C_{ \phi  \chi  \chi}.
\label{eq:C_q}
\end{align}
The expression of the effective $\phi \bar{\chi} \chi$ coupling coefficient ($\phi  =  h, H$), $C_{\phi  \chi  \chi}$, is as follows: 
\begin{align}
             C_{ \phi  \chi  \chi }      =    \frac{ -  m_\chi }{ ( 4  \pi )^2 }  
                                                     \Biggl\{
                                                         &  g_{ \phi  a  a }  (  \xi^\chi_a )^2    \left[  \frac{ \del }{ \del  p^2 }  B_0  ( p^2, m_a^2, m_\chi^2 )\right]_{ p^2  =  m_\chi^2 }
                                                         \nn   \\
                                                         &
                                                         +  g_{ \phi  A  A }  (  \xi^\chi_A )^2    \left[  \frac{ \del }{ \del  p^2 }  B_0  ( p^2, m_A^2, m_\chi^2 )\right]_{ p^2  =  m_\chi^2 }
                                                         \nn   \\
                                                         &
                                                         +  \frac{2  g_{ \phi  a  A }  \xi^\chi_A  \xi^\chi_a}{ m_A^2  -m_a^2 }   \left[ B_1  ( m_\chi^2,  m_A^2,  m_\chi^2 )  -B_1  ( m_\chi^2,  m_a^2,  m_\chi^2 ) \right]    
                                                     \Biggr\}.
                                                     \label{eq:chxx}
\end{align}
The definitions of the loop functions $B_0$ and $B_1$ are given in Appendix \ref{sec:loopfun_b}.

Next, we calculate the triangle diagrams which contribute to the effective DM-gluon coupling, $C_G$. 
There is the relation between $\bar{\chi}  \chi  G^a_{ \mu  \nu }  G^{ a  \mu  \nu }$ and $\bar{\chi}  \chi  \bar{Q}  Q$ via~\cite{Shifman:1978zn}
\begin{align}
            m_Q  \bar{Q}  Q  =  -  \frac{ \alpha_s }{ 12  \pi }  G^a_{ \mu  \nu }  G^{ a  \mu  \nu },
            \label{eq:qqtogg}
\end{align} 
where $Q$ indicates the heavy quark ($c$, $b$, $t$).
Using this relation, $C_G^{\rm tri}$ can be expressed with $C_Q^{\rm tri}$ as follows: 
\begin{align}
 C_{G}^{ \rm tri }  = \sum_{Q=c,b,t}\frac{ 2 }{ 27 }  C_{Q}^{ \rm tri }, 
\end{align}
where $C_Q^{\rm tri}$ is obtained by substituting $q$ to $Q$ in Eq.~(\ref{eq:C_q}).

\subsection{Box diagrams}
\label{sec:dd_box}
In the following, we show the effective operators from the box diagrams.

\label{sec:eo_box}
\subsubsection{DM-quark scalar operators from box diagrams}
\label{sec:boxlq}
We derive the contributions to $C_q$, $C_q^{(1)}$, and $C_q^{(2)}$ in Eq.~\eqref{eq:eff_ope} from the box diagrams shown in Fig.~\ref{fig:boxcomb}. 
Because the quark in the external line is non-relativistic, we expand the amplitude by the external quark momentum and derive the effective operators. 
After that, we decompose these effective operators into the scalar and twist-2 operators as follows: 
        \begin{figure}[]
          \begin{center}
            \begin{tabular}{c}
              \begin{minipage}{0.4\hsize}
                \begin{center}
                \includegraphics[width=5cm]{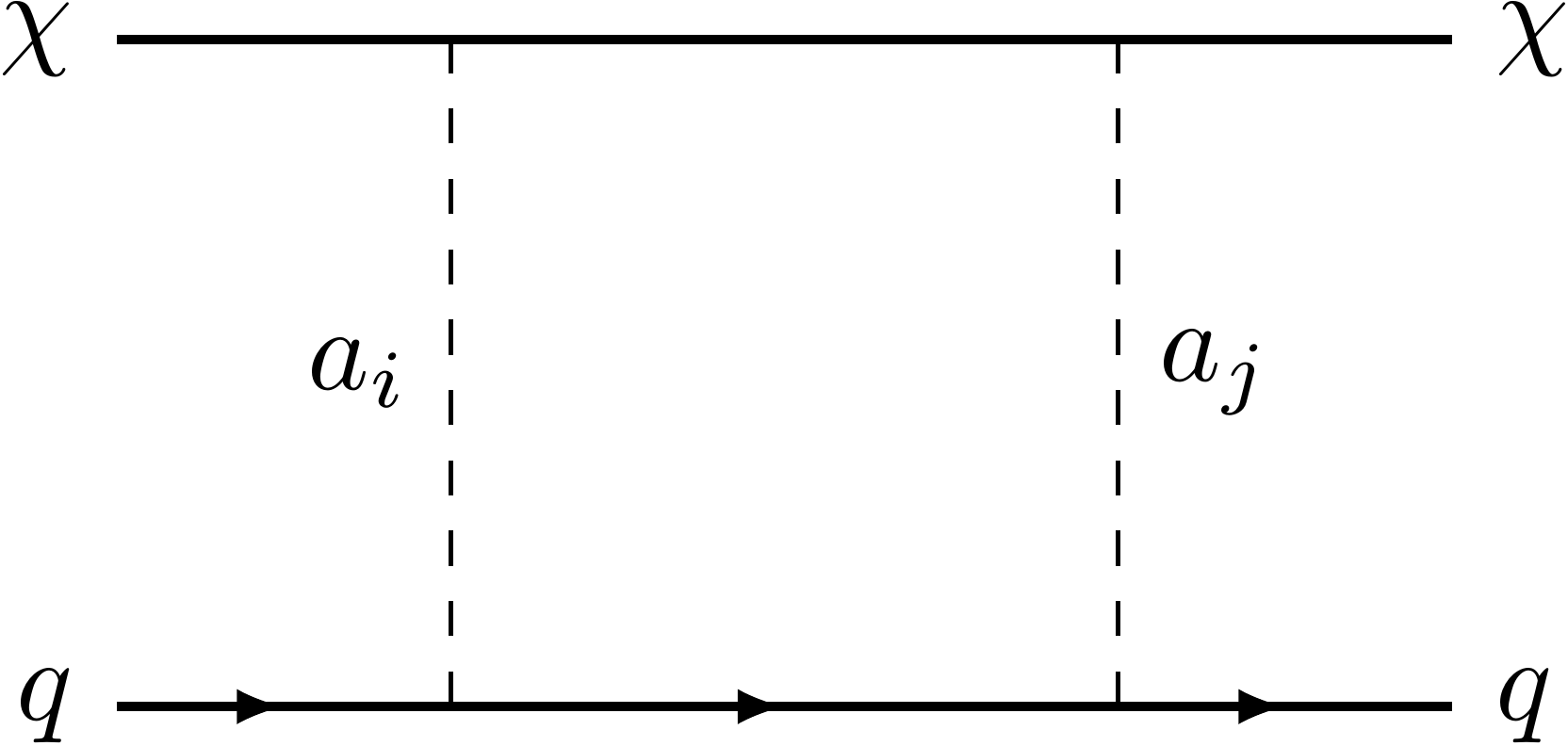}
                \end{center}
              \end{minipage}
              \begin{minipage}{0.4\hsize}
                \begin{center}
                \includegraphics[width=5cm]{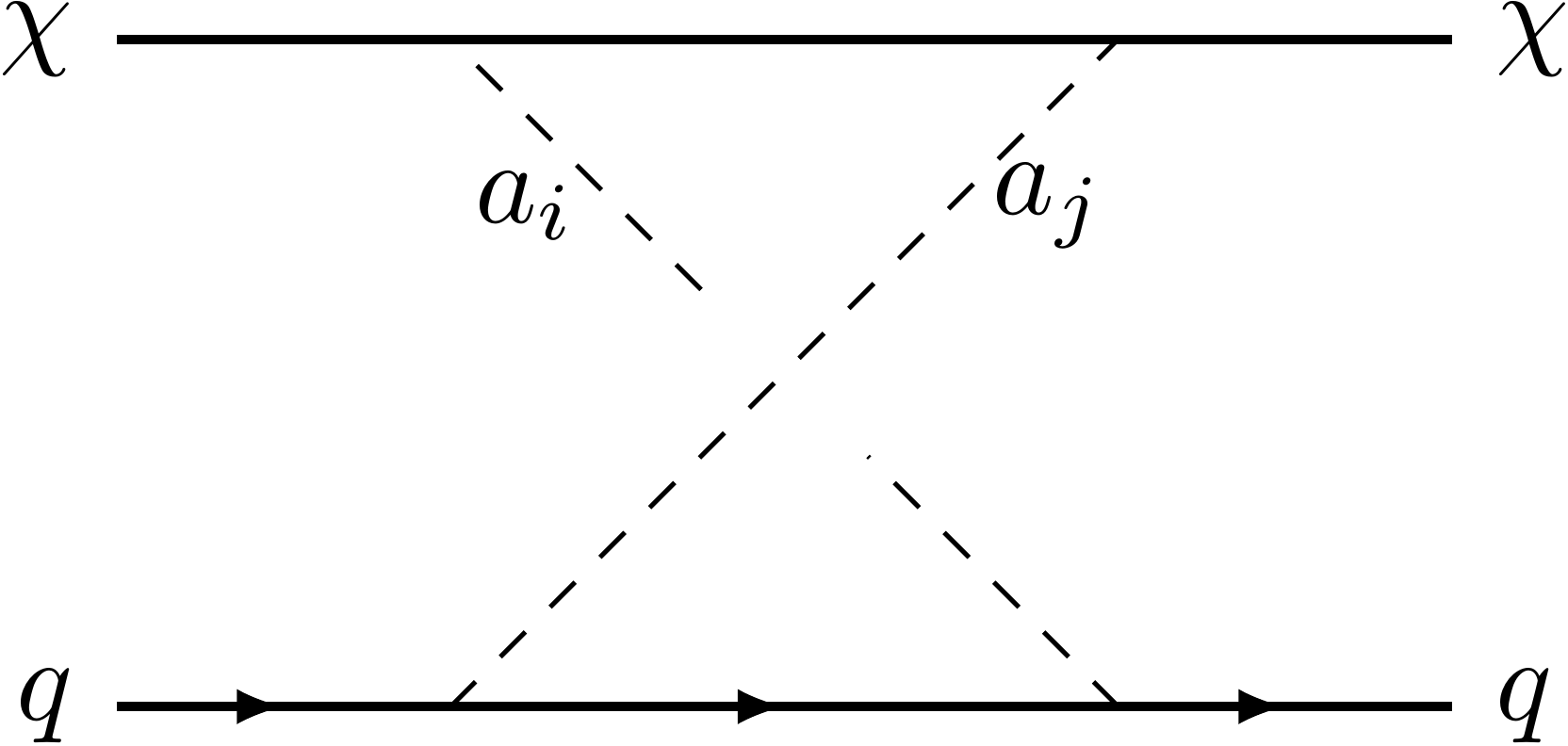}
                \end{center}
              \end{minipage}
            \end{tabular}
          \end{center}
          \caption{The box diagrams which induce the DM-quark effective operators, where $a_i  =  a, A$. }
          \label{fig:boxcomb}
        \end{figure}
        \begin{align}
            { \cal  L }_{ \rm  eff }    \supset&    \frac{ 1 }{ 2 }        C_q^{ \rm box }                 m_q  \bar{ \chi }      \chi  \bar{ q }  q
                                                           +    \frac{ 1 }{ 2 }
                                                                     \left[
                                                                             C_q^{ ( 1 )  { \rm box } }    \bar{ \chi }  i  \del^\mu  \gamma^\nu  \chi  { \cal  O }_{ \mu  \nu }^q
                                                                         +  C_q^{ ( 2 )  { \rm box } }    \bar{ \chi }  i  \del^\mu  i  \del^\nu  \chi       { \cal  O }_{ \mu  \nu }^q
                                                                     \right].
        \end{align}
        The Wilson coefficients are
        \begin{align}
            C_q^{ \rm box }                 
                 =    \frac{ -  m_\chi }{ ( 4  \pi )^2 }  \left( \frac{ m_q }{ v } \right)^2
                        \Biggl\{ 
                        &
                            \frac{ ( \xi^\chi_a  \xi^q_a )^2 }{ m_a^2 }  \left[ G ( \scalebox{0.92}{$\displaystyle  m_\chi^2,  0,  m_a^2  $} )  -  G ( \scalebox{0.92}{$\displaystyle  m_\chi^2,  m_a^2,  0 $} ) \right]
                            \nn   \\
                            &  +
                            \frac{ ( \xi^\chi_A  \xi^q_A )^2 }{ m_A^2 }  \left[ G ( \scalebox{0.92}{$\displaystyle  m_\chi^2,  0,  m_A^2 $} )  -  G ( \scalebox{0.92}{$\displaystyle  m_\chi^2,  m_A^2,  0 $} ) \right]
                            \nn   \\
                            &  +
                            2  \frac{ \xi^\chi_A  \xi^\chi_a  \xi^q_A  \xi^q_a }{ m_A^2  -  m_a^2 }  \left[ G ( \scalebox{0.92}{$\displaystyle  m_\chi^2,  m_A^2,  0 $} )  -  G ( \scalebox{0.92}{$\displaystyle  m_\chi^2,  m_a^2,  0 $} ) \right]
                        \Biggr\},
                        \label{eq:scalarwc}
                        \\
                        \nn   \\
            C_q^{ ( 1 )  { \rm box } }     
                 =    \frac{ -  8 }{ ( 4  \pi )^2 }  \left( \frac{ m_q }{ v } \right)^2
                        \Biggl\{ 
                        &
                            \frac{ ( \xi^\chi_a  \xi^q_a )^2 }{ m_a^2 }  \left[ X_{001} ( \scalebox{0.92}{$\displaystyle  p^2,  m_\chi^2,  0,  m_a^2 $} )  -  X_{001} ( \scalebox{0.92}{$\displaystyle  p^2,  m_\chi^2,  m_a^2,  0 $} ) \right]
                            \nn   \\
                            &  +
                            \frac{ ( \xi^\chi_A  \xi^q_A )^2 }{ m_A^2 }  \left[ X_{001} ( \scalebox{0.92}{$\displaystyle  p^2,  m_\chi^2,  0,  m_A^2 $} )  -  X_{001} ( \scalebox{0.92}{$\displaystyle  p^2,  m_\chi^2,  m_A^2,  0 $} ) \right]
                            \nn   \\
                            &  +
                            2  \frac{ \xi^\chi_A  \xi^\chi_a  \xi^q_A  \xi^q_a }{ m_A^2  -  m_a^2 }  \left[ X_{001} ( \scalebox{0.92}{$\displaystyle  p^2,  m_\chi^2,  m_A^2,  0 $} )  -  X_{001} ( \scalebox{0.92}{$\displaystyle  p^2,  m_\chi^2,  m_a^2,  0  $} ) \right]
                        \Biggr\},
                        \label{eq:twist1wc}
                        \\
                        \nn   \\
            C_q^{ ( 2 )  { \rm box } }
                 =    \frac{ -  4  m_\chi }{ ( 4  \pi )^2 }  \left( \frac{ m_q }{ v } \right)^2
                        \Biggl\{ 
                        &
                            \frac{ ( \xi^\chi_a  \xi^q_a )^2 }{ m_a^2 }  \left[ X_{111} ( \scalebox{0.92}{$\displaystyle  p^2,  m_\chi^2,  0,  m_a^2 $} )  -  X_{111} ( \scalebox{0.92}{$\displaystyle  p^2,  m_\chi^2,  m_a^2,  0 $} ) \right]
                            \nn   \\
                            &  +
                            \frac{ ( \xi^\chi_A  \xi^q_A )^2 }{ m_A^2 }  \left[ X_{111} ( \scalebox{0.92}{$\displaystyle  p^2,  m_\chi^2,  0,  m_A^2 $} )  -  X_{111} ( \scalebox{0.92}{$\displaystyle  p^2,  m_\chi^2,  m_A^2,  0 $} ) \right]
                            \nn   \\
                            &  +
                            2  \frac{ \xi^\chi_A  \xi^\chi_a  \xi^q_A  \xi^q_a }{ m_A^2  -  m_a^2 }  \left[ X_{111} ( \scalebox{0.92}{$\displaystyle  p^2,  m_\chi^2,  m_A^2,  0 $} )  -  X_{111} ( \scalebox{0.92}{$\displaystyle  p^2,  m_\chi^2,  m_a^2,  0 $} ) \right]
                        \Biggr\},
                        \label{eq:twist2wc}
        \end{align}
        where
        \begin{align}
            G  ( m_\chi^2,  m_1^2,  m_2^2 )    =    6  X_{001} ( m_\chi^2,  m_\chi^2,  m_1^2,  m_2^2 )  +  m_\chi^2  X_{111} ( m_\chi^2,  m_\chi^2,  m_1^2,  m_2^2 ).
        \end{align}
        The definitions of the loop functions $X_{001}$ and $X_{111}$ are given in Appendix \ref{sec:loopfun_x}.
        The details of the derivation of the coefficients in Eqs.~(\ref{eq:scalarwc})--(\ref{eq:twist2wc}) are given in Appendix \ref{sec:box_cq_detail}.

        \subsubsection{DM-gluon scalar operator from box diagrams}
        \label{sec:box_glue}
  \begin{figure}[H]
    \begin{minipage}{1\hsize}
        \begin{center}
          \begin{minipage}{0.25\hsize}
            \begin{center}
            \includegraphics[width=3.5cm]{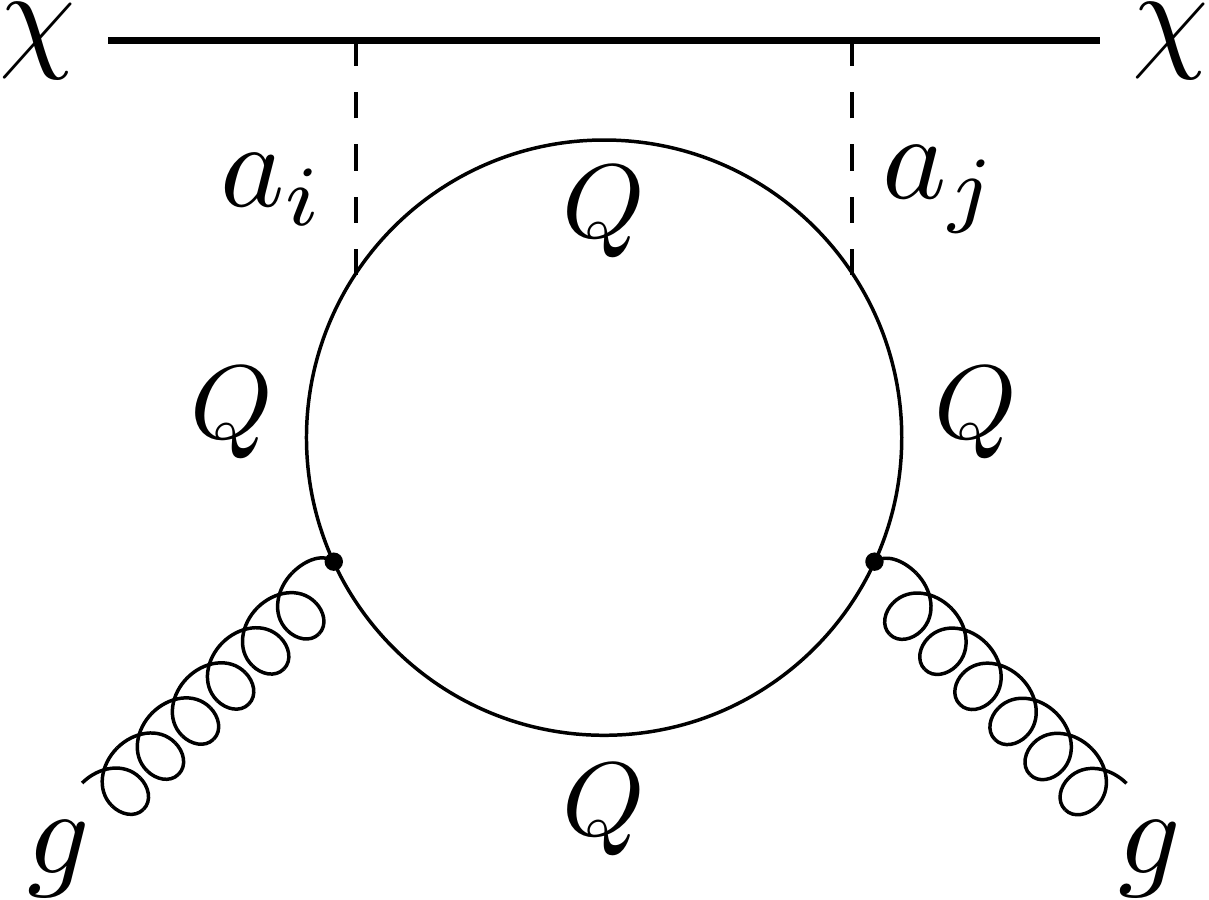}
            \end{center}
          \end{minipage}
          $+$
          \begin{minipage}{0.25\hsize}
            \begin{center}
            \includegraphics[width=3.5cm]{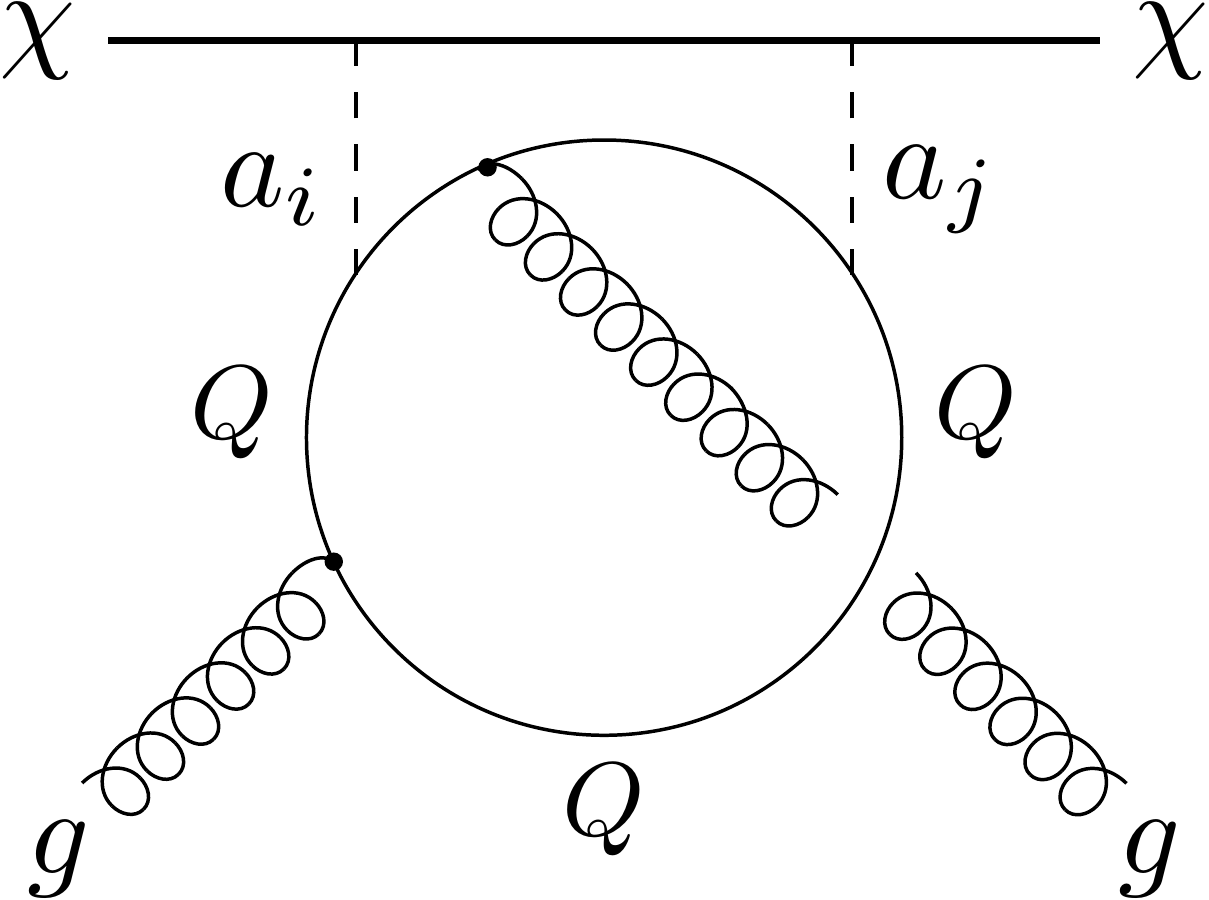}
            \end{center}
          \end{minipage}
          \hspace{0.5cm}
          $\to$
          \begin{minipage}{0.3\hsize}
            \begin{center}
                \includegraphics[width=2.8cm]{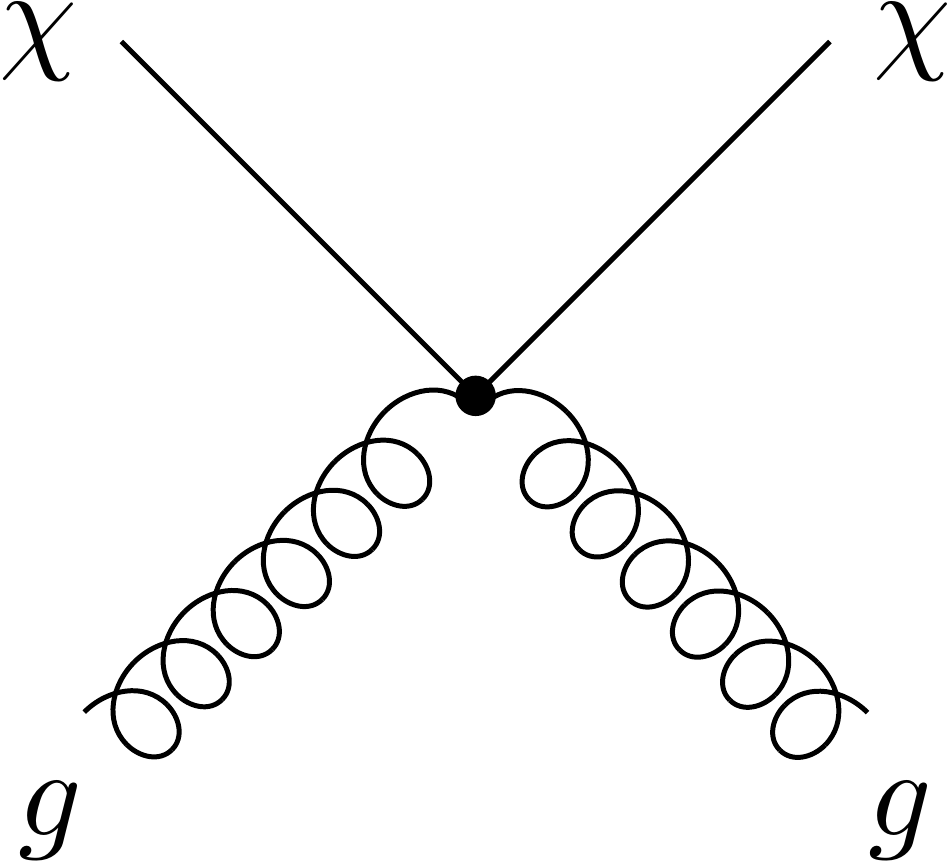}
                \end{center}
              \end{minipage}
          \end{center}
          \end{minipage}
          \\
          \\
          \caption{The two-loop diagrams for the DM-gluon effective operator, where $Q = c, b, t$. 
                        }
          \label{fig:triboxcomp}
        \end{figure}
We calculate the contribution from the box diagrams to $C_G$. 
For the box diagrams, the procedure of the triangle diagrams cannot be applied due to the following reasons: 
First, for $m_Q  >  m_\chi, m_{a}, m_{A}$, we cannot obtain the effective operators with the heavy quark $\bar{\chi}  \chi  \bar{Q}  Q$ by expanding the amplitude by the quark momentum as done in Sec.~\ref{sec:boxlq}.  
In particular, the loop calculation is mandatory for $m_t  > m_a$ as we will see in later. 
Second, even if $m_Q  \ll  m_\chi, m_{a}, m_{A}$, 
the second diagrams in Fig.~\ref{fig:triboxcomp} are not included 
if we derive $C_G^{\rm box}$ from $C_Q^{\rm box}$ by using Eq.~(\ref{eq:qqtogg}). 
Thus, it is necessary to calculate the two-loop diagrams shown in Fig.~\ref{fig:triboxcomp} 
and to read out the effective operator $\bar{\chi}  \chi  G^a_{ \mu \nu }  G^{ a \mu \nu }$ directly.  
We use the Fock-Schwinger gauge for the gluon field~\cite{Novikov:1983gd}. 
This gauge enables us to calculate the effective operator much more transparently~\cite{1007.2601}. 
In the end, we find the following effective operator from these two-loop diagrams~\cite{1501.04161}.  
        \begin{align}
            { \cal  L }_{\rm  eff}    &\supset    \frac{ 1 }{ 2 }  C_G^{\rm box}  \left( \frac{ -9  \alpha_s }{ 8  \pi }  \ol{\chi}  \chi  G^a_{ \mu \nu }  G^{ a \mu \nu } \right).
            \label{eq:xxgg_box}
        \end{align}
        The Wilson coefficient is given by
        \begin{align}
            C_G^{\rm box}      =    \sum_{Q  =  c,  b,  t}  \frac{ -  m_\chi }{ 432  \pi^2 }
                                                \left( \frac{m_Q}{v} \right)^2
                                                \Biggl[
                                                  ( \xi^{\chi}_a  \xi^Q_a )^2  \frac{ \del  F( m_a^2 ) }{ \del  m_a^2 }
                                                  &
                                              +  ( \xi^{\chi}_A  \xi^Q_A )^2  \frac{ \del  F( m_A^2 ) }{ \del  m_A^2 }
                                              \nn   \\
                                              &
                                              +  2  \xi^{\chi}_A  \xi^{\chi}_a  \xi^Q_A  \xi^Q_a  \frac{ \left[ F  ( m^2_A )  -  F  ( m^2_a ) \right] }{ m_A^2  -  m_a^2 }
                                                \Biggr],
                                                \label{eq:cg_box}
        \end{align}
        where
        \begin{align}
            F  ( m_a^2 )      =    \int_0^1  dx
                                            \biggl\{
                                                          &
                                                           3  Y_1  ( p^2,  m_\chi^2,  m_a^2,  m_Q^2 )  
                                                           \nn   \\
                                                          &  -  m_Q^2  \frac{ ( 2  +  5  x  -  5  x^2 ) }{ x^2 ( 1  -  x )^2 }  Y_2  ( p^2,  m_\chi^2,  m_a^2,  m_Q^2 )  
                                                          \nn   \\
                                                          &  -  2  m_Q^4  \frac{ ( 1  -  2  x  +  2  x^2 ) }{ x^3 ( 1  -  x )^3 }  Y_3  ( p^2,  m_\chi^2,  m_a^2,  m_Q^2 )
                                            \biggr\}.
        \end{align}
        The definitions of the loop functions $Y_1$,  $Y_2$, and $Y_3$ are shown in Appendix \ref{sec:loopfun_y}.
        The expression for $ \del  F( m_a^2 )  /  \del  m_a^2 $ is shown in Appendix \ref{sec:dfdm2}.        
        The details of the derivation of $C_G^{\rm box}$ are given in Appendix \ref{sec:box_2loop_detail}.

\section{Numerical Analysis}
\label{sec:result}
In this section, we show our numerical analysis for the DM-nucleon SI scattering cross section $(\sigma_{\text{SI}})$.
We focus on the region of the parameter space where the DM thermal relic abundance matches the measured value of the DM energy density,
$\Omega h^2 = 0.1198 \pm 0.0015$~\cite{1502.01589}. 
In Sec.~\ref{sec:annihilation}, we show that
it is easy to realize the correct DM energy density by choosing $g_\chi$ appropriately. 
Using the value of $g_\chi$, we calculate $\sigma_{\text{SI}}$.
In Sec.~\ref{sec:vsOtherWorks}, 
we show the comparison of our result with the previous one~\cite{1711.02110}. 
We find that the gluon contribution through the box diagrams was drastically changed from the results in~\cite{1711.02110}. 
In Sec.~\ref{sec:future}, we discuss the effect of the scalar quartic coupling which enhances $\sigma_{\rm  SI}$. 
We find that some parameter points are within the reach of the XENONnT~\cite{1512.07501} and LZ experiments~\cite{1611.05525}.

\subsection{Determination of $g_\chi$ through the DM thermal relic abundance}
\label{sec:annihilation}
    \begin{figure}[H]
        \centering
        \begin{minipage}{0.4\hsize}
            \begin{center} 
                \includegraphics[width=4cm]{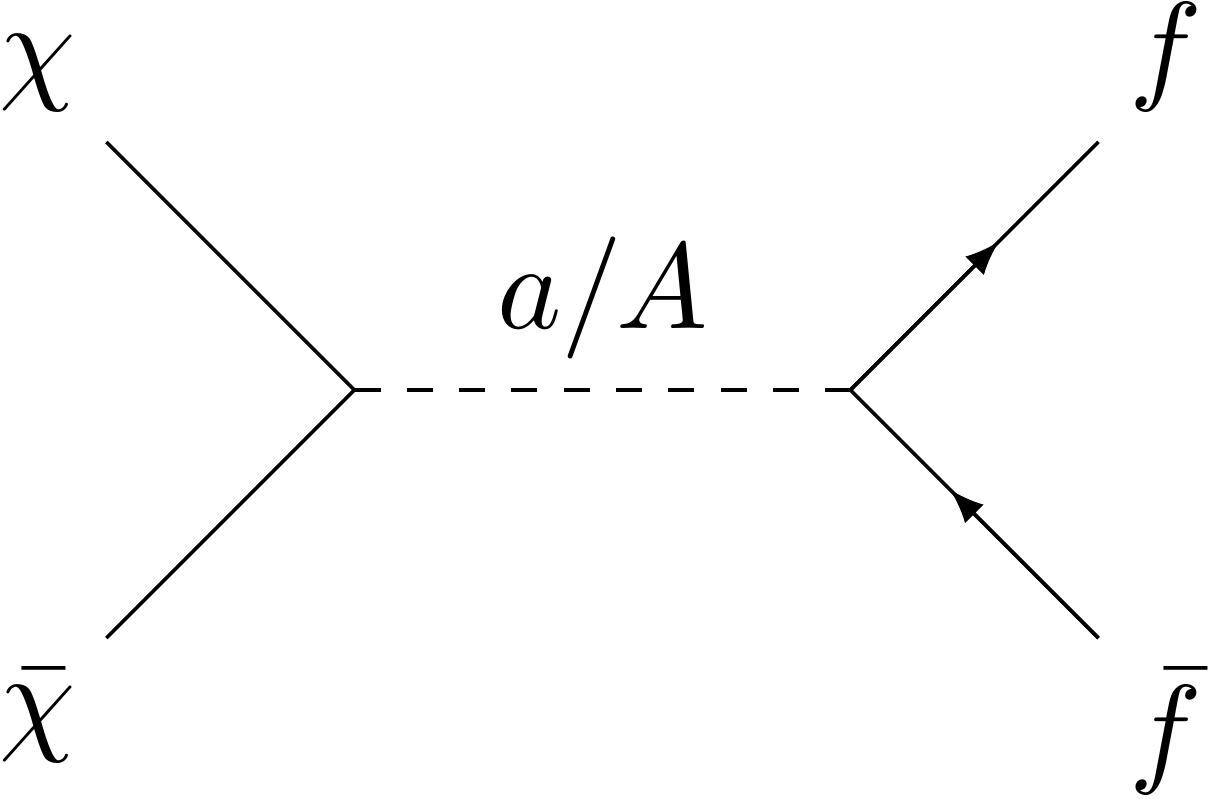}
            \end{center}
        \end{minipage}
        \begin{minipage}{0.4\hsize}
            \begin{center} 
                \includegraphics[width=4cm]{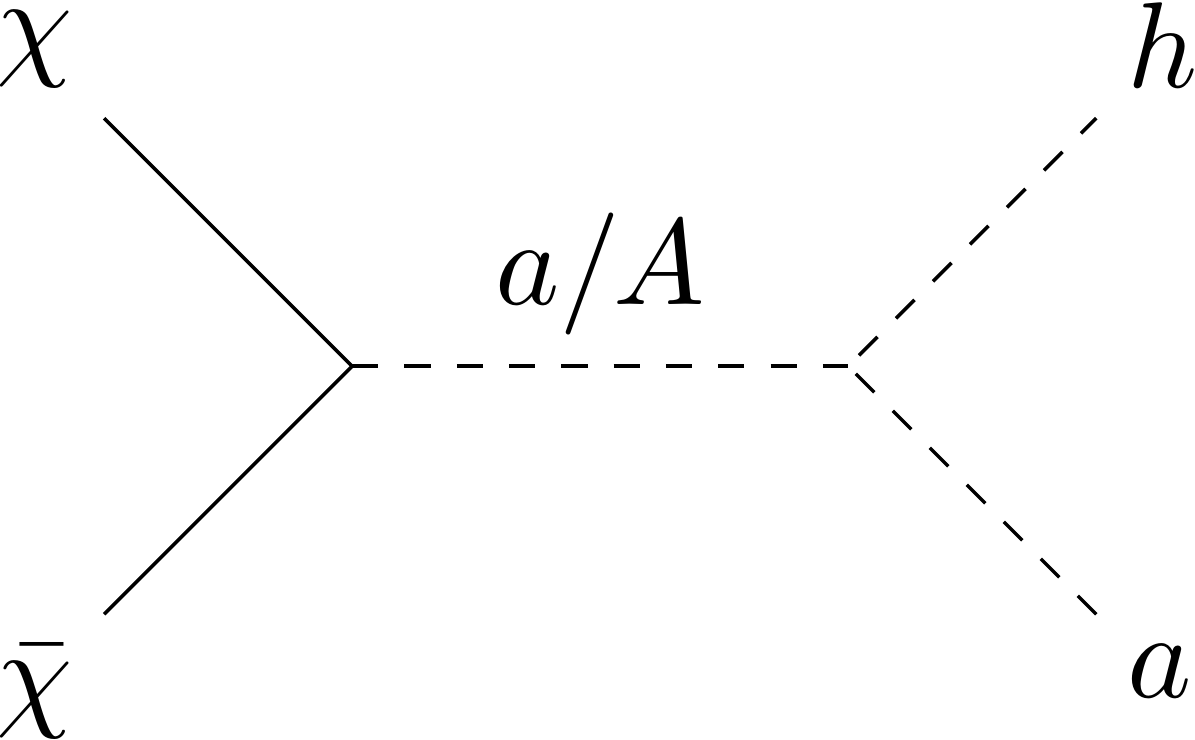}
            \end{center}
        \end{minipage}
        \caption{The dominant annihilation diagrams at the tree-level, where $f$ indicates the SM fermion. }
        \label{fig:relic}
    \end{figure}
We discuss the DM pair annihilation and determine $g_\chi$ to realize the measured value of the DM energy density by the thermal relic abundance.
The dominant annihilation processes are shown in Fig.~\ref{fig:relic}. 
Note that the $f  \bar{f}$ channel, where $f$ is the SM fermion, depends on the THDM-types
through the rescaling factors of the Yukawa couplings, $\xi^f_a$ and $\xi^f_A$. 
On the other hand, the scalar channel is independent of the THDM-type.

        \begin{figure}[]
        \centering
            \begin{tabular}{c}
            \centering
                \begin{minipage}{0.4\hsize}
                    \includegraphics[width=6.6cm]{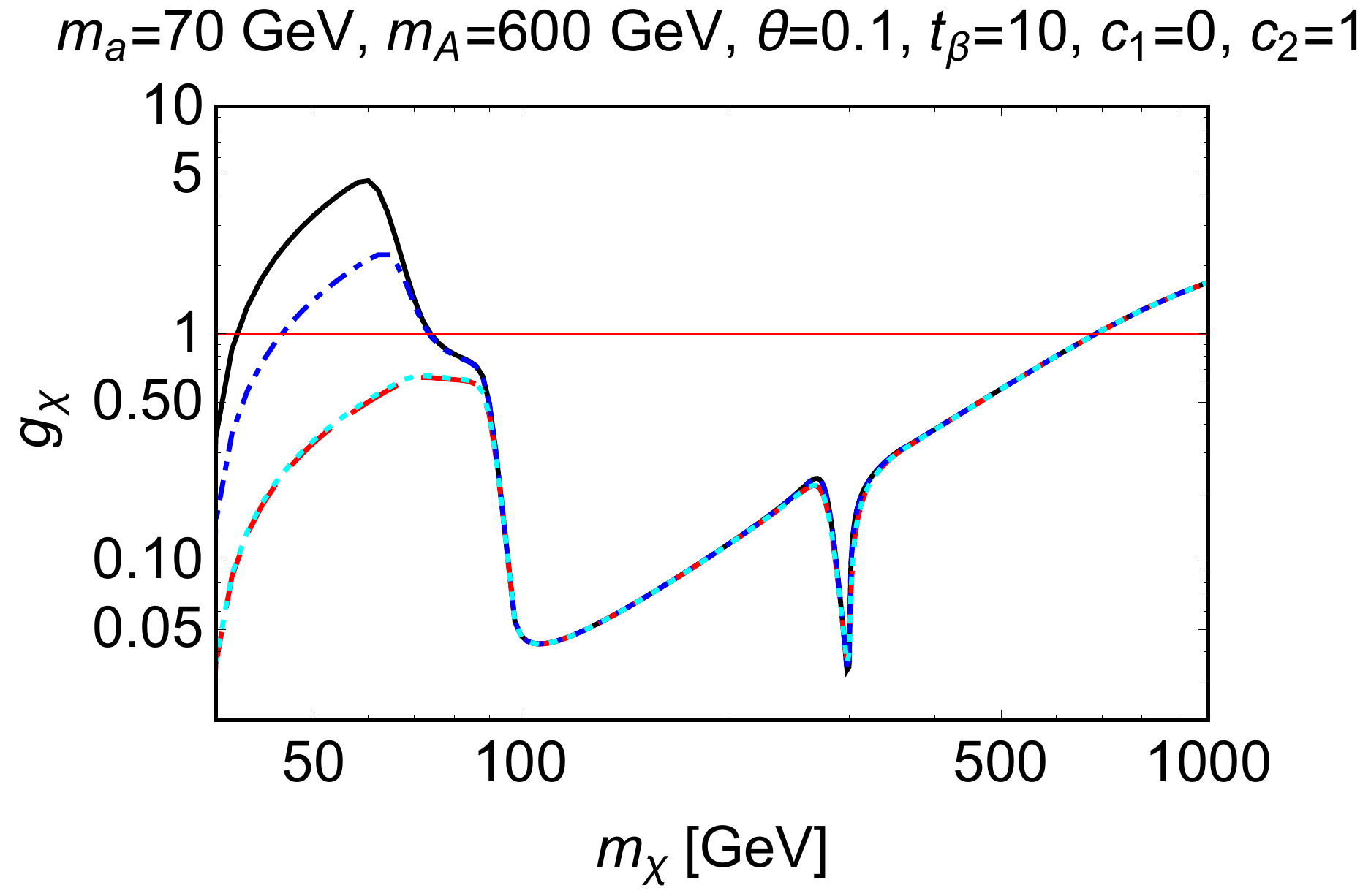}
                \end{minipage}
                ~~
                \begin{minipage}{0.6\hsize}
                    \includegraphics[width=9cm]{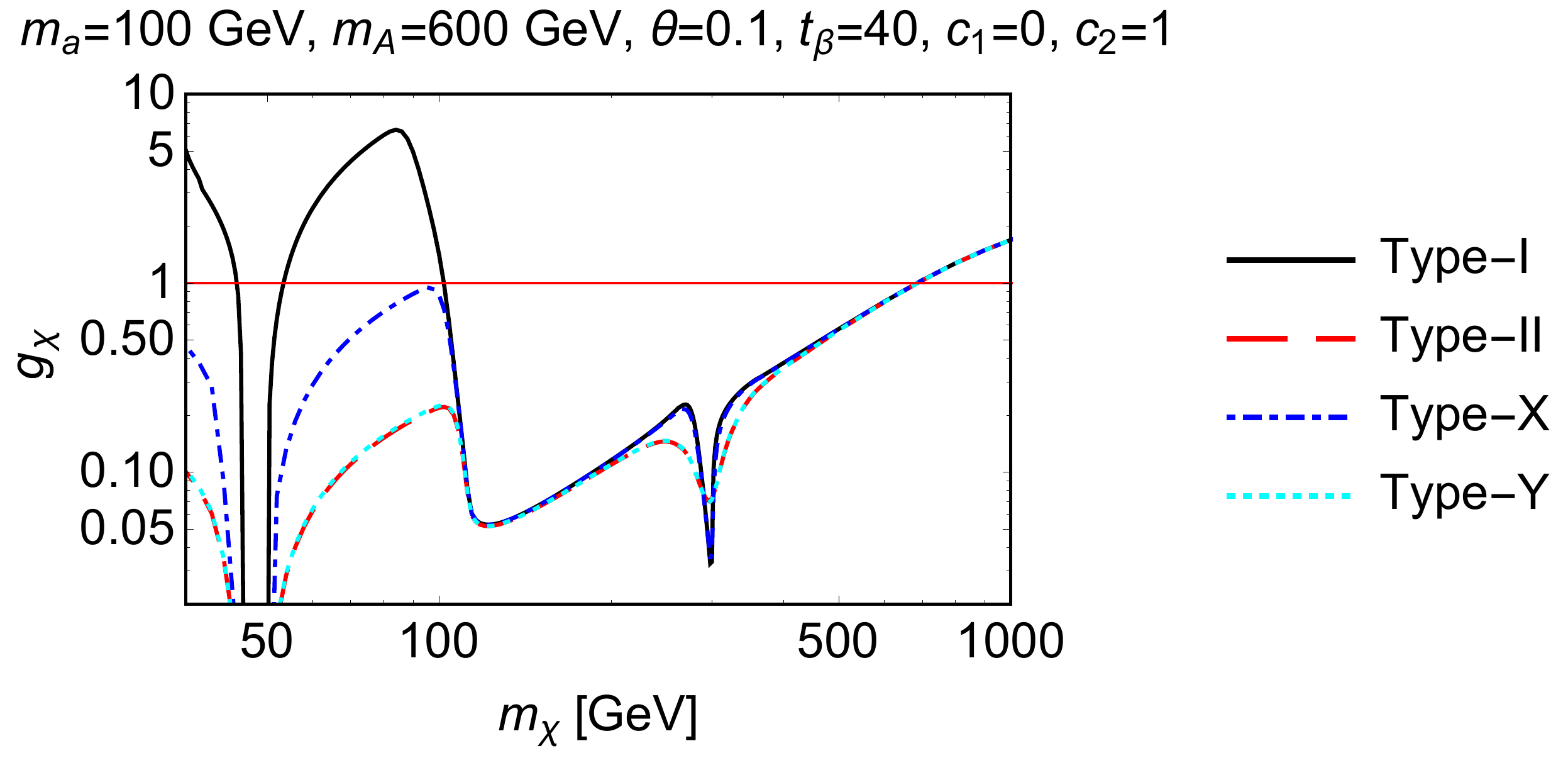}
                \end{minipage}
            \end{tabular}
            \caption{The DM-pseudoscalar coupling $g_\chi$ as a function of $m_\chi$.  
                     The left panel shows $g_\chi$ for $m_a  =  70$~GeV and $t_\beta=10$, 
	             and the right for $m_a  =  100$~GeV and $t_\beta=40$. 
                     The other parameters are $m_A = 600$ GeV, $\theta = 0.1$, $c_1 = 0$, $c_2 = 1$ for both of the panels. 
                     The black solid, red dashed, blue dot-dashed, and cyan dotted lines show the type-I, II, X, and Y, respectively. 
                     The red horizontal line indicates $g_\chi  =  1$. 
                     The coupling of the type-Y turns out to be almost the same result as that of the type-II. 
                      }
            \label{fig:gDM_type_demo}
        \end{figure}
Figure~\ref{fig:gDM_type_demo} shows $g_\chi$ as a function of $m_\chi$ for the four THDM-types.
Here we use \texttt{micrOMEGAs}~\cite{1305.0237} to calculate the DM thermal relic abundance
for the determination of $g_\chi$.
We find that $g_\chi$ becomes suddenly small around the funnel position where $m_\chi  \sim  m_{a_i}/2 \ ( a_i  =  a, A )$. 
In these regions, 
the s-channel amplitude becomes very large
because it is proportional  to $(s  -  m_{a_i}^2 + i m_{a_i} \Gamma_{a_i})^{-1}$,
where $s$ is the invariant mass square and $\Gamma_{a_i}$ is the decay width of $a_i$. 
As a result, $g_\chi$ has to be small to obtain $\Omega h^2 \sim 0.12$,
otherwise the relic abundance becomes too small. 
The coupling also becomes small after the new annihilation channel $\chi \bar{\chi}  \to  h  a$ opens. 
As can be seen in Fig.~\ref{fig:gDM_type_demo}, $g_\chi$ suddenly begins to decrease at $m_\chi  \sim  ( m_h  +  m_a )/2$. 
For the larger $m_\chi$, the annihilation amplitude is suppressed by $m_\chi^{-2}$, 
and $g_\chi$ increases in proportion to $m_\chi^2$. 
We find that $g_\chi > 1$ for $m_\chi  \geq  690$~GeV.

Figure~\ref{fig:gDM_type_demo} also shows the THDM-type dependence of $g_\chi$. 
The type dependence appears in the region where the annihilation channel $\chi  \bar{\chi}  \to f\bar{f}$ is dominant.
For $m_\chi \leq (m_h + m_a)/2$, the channel $\chi \bar{\chi}  \to  h  a$ is kinematically forbidden, 
and thus the type dependence appears in $g_\chi$. 
For the type-I, all of the Yukawa couplings are suppressed by large $t_\beta$, and $g_\chi$ tends to be large to keep $\Omega h^2  \sim  0.12$. 
For the type-II and the type-Y, $g_\chi$ is almost the same. 
This is because the difference in the charged lepton sector is negligible if the down-type quark Yukawa couplings have $t_\beta$ enhancement.
The annihilation channel $\chi \bar{\chi}  \to  h  a$ dominates the process once allowed kinematically, 
and thus $g_\chi$ becomes type independent for $m_\chi > (m_h + m_a)/2$.
Around $m_\chi  \sim  m_A/2$, however, we find the type dependence of $g_\chi$ again. 
This is because the annihilation channel to $f \bar{f}$ through the mediator $A$ again dominates 
the annihilation process by $t_\beta$ enhancement. 
In Table~\ref{tab:type_annihilation_channel}, we show the dominant annihilation channels near the funnel positions for each of the THDM-types. 
\begin{table}[H]
\centering
\begin{tabular}{|  l  || c  | c  |}\hline
 & $m_\chi  \sim  50$ GeV & $m_\chi  \sim  300$ GeV 
 \\\hline \hline
Type-I  &  \scalebox{0.85}{ $b  \bar{b}$ }  &  \scalebox{0.85}{ $h  a$ }
\\\hline
Type-II  &  \scalebox{0.85}{ $ b  \bar{b} $ }  &  \scalebox{0.85}{ $b  \bar{b}$ }
\\\hline
Type-X  &  \scalebox{0.85}{ $ \tau^-  \tau^+ $ }  &  \scalebox{0.85}{ $\tau^-  \tau^+$ }
\\\hline
Type-Y  &  \scalebox{0.85}{ $ b  \bar{b} $ }  &  \scalebox{0.85}{ $b  \bar{b}$ }
\\\hline
\end{tabular}
\caption{The dominant annihilation channels of each THDM-type near the funnel positions. 
              The parameters are $m_a = 100$ GeV, $m_A = 600$ GeV, $\theta = 0.1$, $t_\beta=40$, $c_1 = 0$, $c_2 = 1$.
              }
\label{tab:type_annihilation_channel}
\end{table}

\subsection{Comparison with the previous results}
\label{sec:vsOtherWorks}

In the following, we compare our result with the previous one in~\cite{1711.02110} at the benchmark point with $m_a  =  100$ GeV, $m_A  =  600$ GeV, $\theta  =  0.1$, $t_\beta  =  40$, $g_\chi  =  1$, 
and the Yukawa structure is the type-II.\footnote{
Note that this combination of the value of $t_\beta$ and $m_A$ with the type-II THDM is already excluded by LHC experiments~\cite{CMS-PAS-HIG-18-005}. 
We take this point only for the comparison.}
We also choose $c_1  =  c_2  =  0$ for the comparison. 
If the heavy scalar masses are degenerated and $c_1  =  c_2  =  0$ under the alignment limit, then $g_{Ha_i a_j} =  0$ where $a_i = a, A$ (see, Appendix \ref{sec:3scalar}). 
We use \texttt{LoopTools}~\cite{Hahn:1998yk} in the numerical calculations of the loop functions. 
Unless otherwise noted, the previous work means~\cite{1711.02110} in this Sec.~\ref{sec:vsOtherWorks}.

\begin{figure}[]
              \centering
                  \begin{minipage}{1\hsize}
                    \centering
                    \includegraphics[width=13cm]{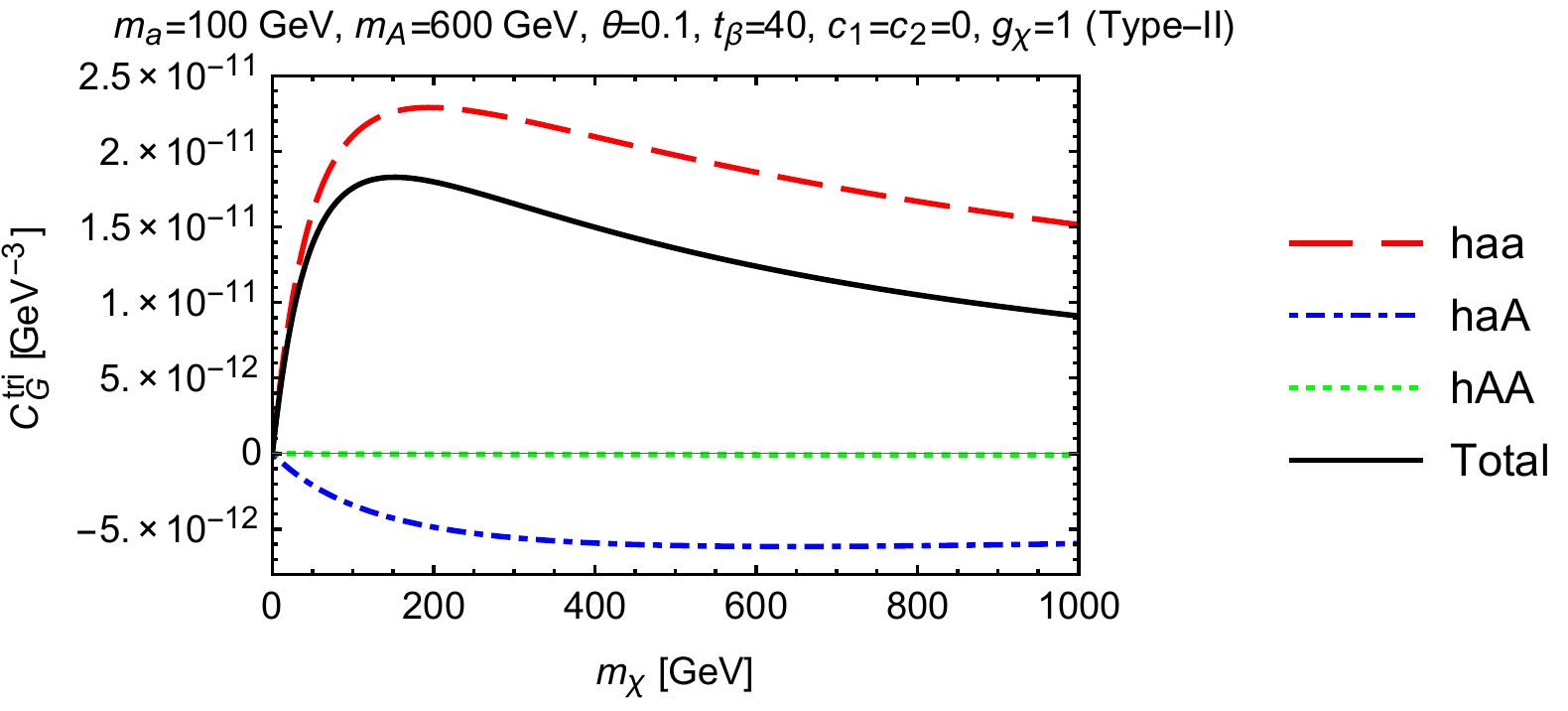}
                    \end{minipage}
                 \caption{The contributions from each of the triangle diagrams to $C_G^{\rm tri}$. 
                               The parameters are $m_a  =  100$ GeV, $m_A  =  600$ GeV, $\theta = 0.1$, $t_\beta  =  40$, $c_1  =  c_2  =  0$, and $g_\chi  =  1$. 
                               The THDM-type is the type-II. 
                               The red dashed, blue dot-dashed, and green dotted lines show the contributions from the $haa$-diagram, the $haA$-diagram, and the $hAA$-diagram, respectively. 
                               The black solid line shows the total of these contributions. 
                               }
             \label{fig:compare_triangle}
    \end{figure}

There are  two improvements in our analysis of the triangle diagrams.
First, we read out the scalar trilinear couplings not only from $V_{\rm port}$ but also from $V_{\rm THDM}$. 
We find that the values of $g_{haa}$ and $g_{haA}$ do not change drastically at the benchmark point. 
On the other hand, the value of $g_{hAA}$ changes largely. 
However, the diagram with $g_{hAA}$ gives the smaller contribution than the diagrams with $g_{haa}$ and $g_{haA}$. 
As a result, the numerical impact of this improvement is negligible. 
Second, 
we include all of the triangle diagrams into our analysis. 
The diagrams with $g_{haA}$ and $g_{hAA}$ were not included in the previous work.
However, these diagrams are also important as pointed out in~\cite{1803.01574}. 
In Fig.~\ref{fig:compare_triangle},  we show the contributions from each of the triangle diagrams to $C_G^{\rm tri}$. 
The red dashed, blue dot-dashed, and green dotted lines are the contributions from the $haa$-diagram, the $haA$-diagram, and the $hAA$-diagram, respectively.
The black solid line shows the total contribution from these three triangle diagrams. 
As can be seen from the figure, the effect from the $haa$-diagram is dominant. 
We also find that the effect from the $haA$-diagram cannot be negligible because $|g_{haa}| <  |g_{haA}|$ at the benchmark point. 
Moreover, the relative sign between the $haa$-diagram and the $haA$-diagram is opposite, and they partially cancel each other. 
At $m_\chi  =  1$~TeV, for example, we find that the total coefficient turns out to be 0.6 times that of the $haa$-diagram. 
Therefore, the contribution from the triangle diagrams is overestimated in the previous work.

\begin{figure}[]
      \begin{center}
        \begin{tabular}{c}
          \begin{minipage}{1\hsize}
            \begin{center}
    \includegraphics[width=16.5cm]{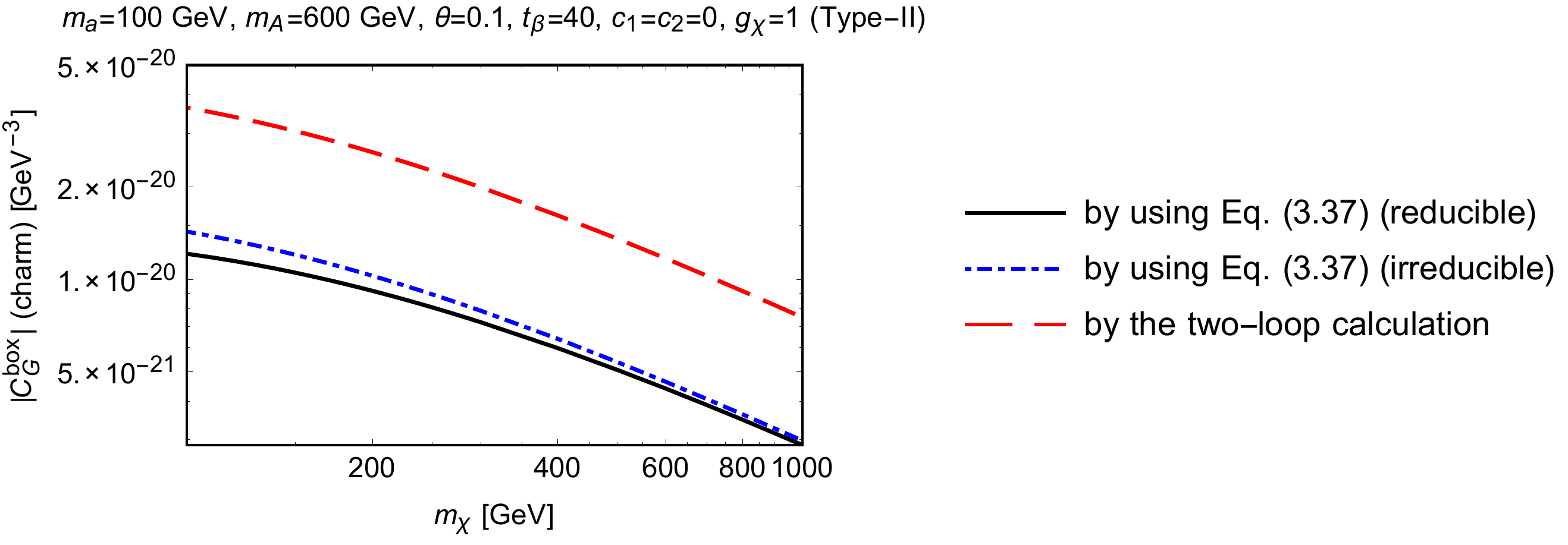}
            \end{center}
          \end{minipage}
          \\
          \\
          \begin{minipage}{1\hsize}
            \begin{center}
    \includegraphics[width=16.5cm]{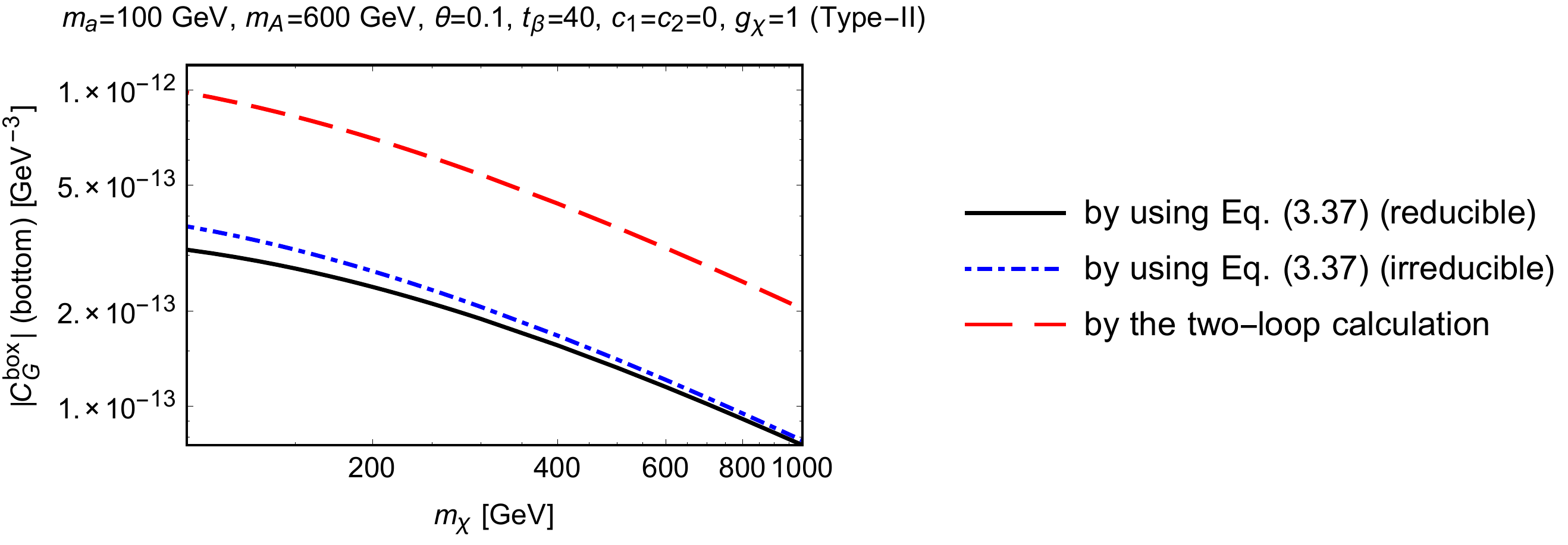}
            \end{center}
          \end{minipage}
          \\
          \\
          \begin{minipage}{1\hsize}
            \begin{center}
    \includegraphics[width=16.5cm]{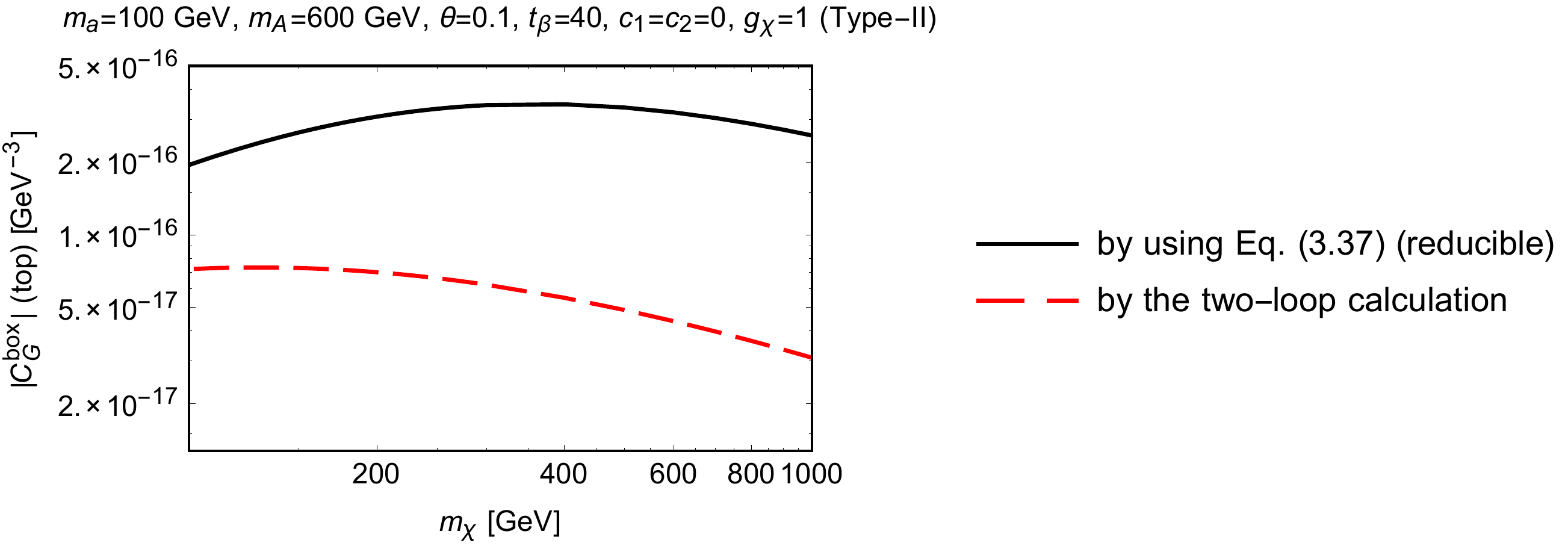}
            \end{center}
          \end{minipage}
        \end{tabular}
        \caption{The contributions from each of the box diagrams to $C_G^{\rm box}$. 
                      The parameters are $m_a  =  100$ GeV, $m_A  =  600$ GeV, $\theta = 0.1$, $t_\beta  =  40$, $c_1  =  c_2  =  0$, and $g_\chi  =  1$. 
                      The THDM-type is the type-II. 
                      The upper, central, and bottom panels show the contributions from the charm, bottom, and top quarks, respectively. 
                      The black solid lines show the contributions which are derived from $C_Q^{\rm  box}$ ($Q  =  c,  b,  t$) without irreducible decomposition by using the relation in Eq.~(\ref{eq:qqtogg}).
                      The blue dotted lines are the same but with the irreducible decomposition. 
                      The red dashed lines show the contributions which are derived from the two-loop calculations.
                      }
        \label{fig:compare_gluon}
      \end{center}
    \end{figure}

As for the box diagrams, there are also two improvements. 
First, as we mentioned in Sec.~\ref{sec:boxlq}, 
we perform the irreducible decomposition into the scalar and twist-2 operators.   
After this decomposition, 
we find new contributions to $C_q^{\rm  box}$ which were not included in the previous work.
See, Appendix \ref{sec:box_cq_detail} for the details. 
We show the numerical impact of the irreducible decomposition in the first two panels in Fig.~\ref{fig:compare_gluon}. 
The black solid line in the upper (central) panel shows $C_G^{\rm  box}$ derived from $C_c^{\rm  box}$ ($C_b^{\rm  box}$) without irreducible decomposition by using the relation in Eq.~(\ref{eq:qqtogg}), which was done in the previous work. 
The blue dotted lines are the same but with the irreducible decomposition. 
We find that the difference between the black and blue lines is small numerically. 
Second, we evaluate $C_G^{\rm  box}$ by calculating the two-loop diagrams shown in Fig.~\ref{fig:triboxcomp}.
The red dashed lines in Fig.~\ref{fig:compare_gluon} show the contributions to $C_G^{\rm  box}$
which are derived by the two-loop calculations. 
As for the contributions from the charm and bottom quarks, we find that $C_G^{\rm  box}$ read out by using Eq.~(\ref{eq:qqtogg}) is $40$\% of the full two-loop calculations. 
Therefore, the previous work underestimated the contributions from the box diagrams with the charm and bottom quarks to $C_G$.
The contribution from the top quark is shown in the last panel in Fig.~\ref{fig:compare_gluon}. 
The black solid line shows $C_G^{\rm  box}$ derived from $C_t^{\rm  box}$ without irreducible decomposition 
by using the relation in Eq.~(\ref{eq:qqtogg}), which was done in the previous work. 
We use Eq.~(A.3) in~\cite{1711.02110} to evaluate $C_t^{\rm box}$.\footnote{
We have found that the overall sign of $C_Q^{\rm box}$ shown in Sec.~\ref{sec:boxlq} disagrees with the previous work. 
In Fig.~\ref{fig:compare_gluon}, we show the comparison with the absolute value of the coefficient. }
The last panel clearly shows that the contribution from the top quark was overestimated in the previous work.  
Thus, it is not justified to relate $C_G^{\rm  box}$ with $C_Q^{\rm  box}$ using Eq.~(\ref{eq:qqtogg}).

As can be seen from Fig.~\ref{fig:compare_gluon}, 
the scattering diagrams with the bottom loop give the dominant contribution to $C_G^{\rm  box}$, 
and thus $C_G^{\rm  box}$ was underestimated in the previous work. 
Comparing Fig.~\ref{fig:compare_triangle} and \ref{fig:compare_gluon}, 
we find that the contribution from the box diagrams is smaller than that from the triangle diagrams in spite of taking the large $t_\beta$ and the type-II THDM.  

\subsection{The DM-nucleon scattering cross section}
\label{sec:future}
We discuss the DM-nucleon scattering cross section numerically. 
In the following, we focus on the triangle diagrams with $h$ 
and search the parameter region where the amplitude of these diagrams becomes large. 
Note that these diagrams are independent of the THDM-types under the alignment limit, 
and thus the cross section is type independent quantitatively. 
As shown in Fig.~\ref{fig:triandbox}, there are also the triangle diagrams with $H$ instead of $h$,
but the amplitude of these diagrams is suppressed by $(m_h/m_H)^2$.

A possible way to enhance $\sigma_{\rm SI}$ is to make $g_{haa}$ large. 
As can be seen from the expression of $g_{haa}$ in Eq.~\eqref{eq:ghaa},
the contributions from $c_1$ and $c_2$ terms to $g_{haa}$  are essential for small $\theta$.
In particular, the $c_2$ term gives a crucial contribution to $g_{haa}$ for the large $t_\beta$ regime. 
Another possibility to enhance $\sigma_{\rm SI}$ is to make $m_a$ as light as possible.  
If $m_a$ is light, the suppression from the loop functions of the triangle diagrams with $a$ is weakened. 
However, we cannot make $m_a$ arbitrary small. 
We find that $m_a  \geq  m_h/2$. 
The constraint on $m_a$ comes from the bound on the branching ratio of the SM Higgs boson.
In the region $m_a  \leq  m_h/2$, this model has a new decay channel of the Higgs boson, $h  \to 2  a$.
The decay width is given by~\cite{1404.3716}
\begin{align}
    \Gamma_{h  \to  2a}  =  \frac{ g_{haa}^2 }{ 32  \pi  m_h }  \sqrt{1  -  \frac{4  m_a^2}{m_h^2}}.
\end{align}
This decay width is proportional to $g_{haa}^2$. 
Note that we are considering the large $g_{haa}$ region to enhance $\sigma_{\rm SI}$. 
Consequently, this decay width becomes large
and gives a strong constraint on $m_a$. 
For example, we find that $\Gamma_{h  \to  2a}  =  4.59$ GeV at the point $m_a  =  60$~GeV,  $m_A  =  600$~GeV, $\theta  =  0.1$, $t_\beta =10$, $c_1  =  0$, and $c_2  =  1$.
This value is much larger than the SM Higgs width. 
The current bound on the Higgs branching ratio into BSM particles is given by the ATLAS experiment~\cite{ATLAS:2018doi},
\begin{align}
   {\rm BR(BSM)}  <  0.26.
\end{align}
The result from the CMS experiment also disfavors the large branching ratio of the Higgs boson into new particles~\cite{CMS-PAS-HIG-17-031}.
We conclude that the parameter region $m_a  \leq  m_h/2$ is excluded.

\begin{figure}[]
        \centering
        \begin{tabular}{c}
        \begin{minipage}{1\hsize}
            \centering
\includegraphics[width=12cm]{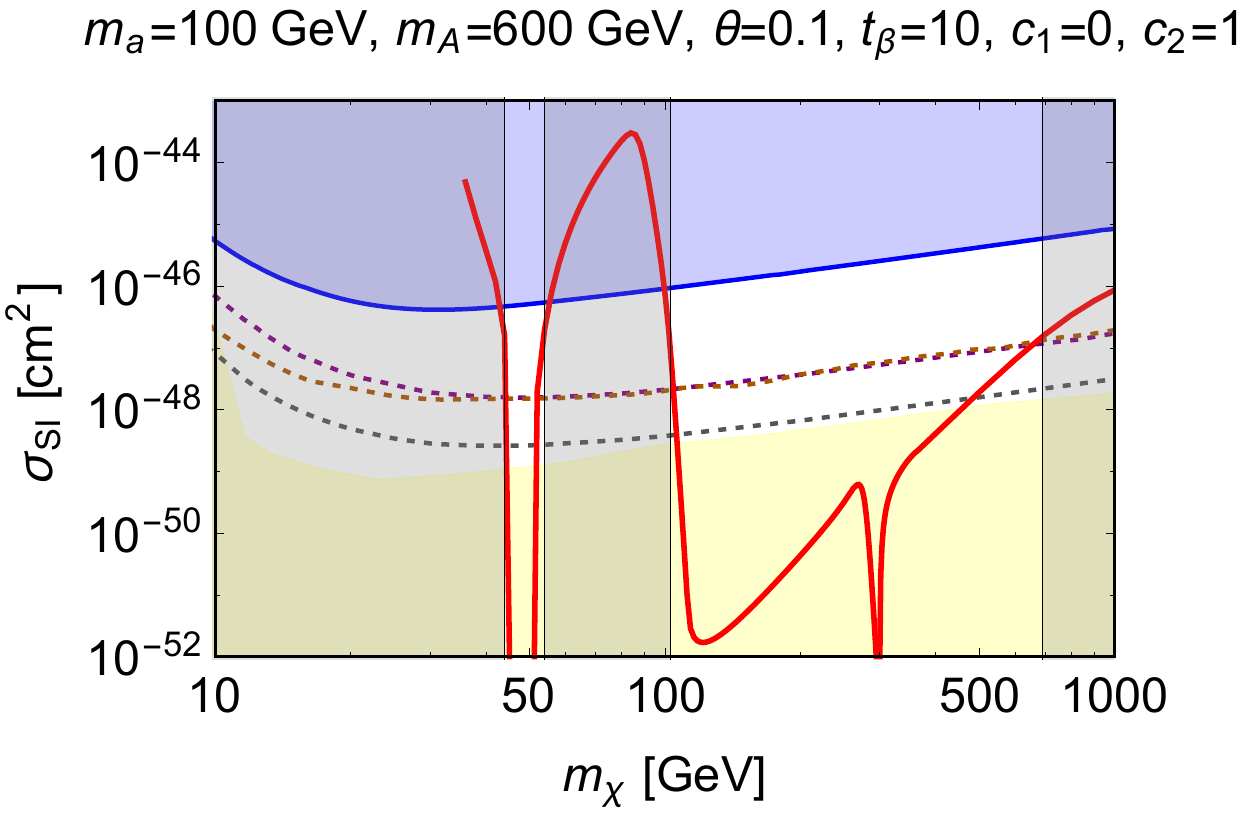}
        \end{minipage}
        \end{tabular}
        \caption{
                      The red solid line shows the predicted SI cross section of this model 
                      with $m_a  =  100$ GeV, $m_A  =  600$ GeV, $\theta = 0.1$, $t_\beta  =  10$, $c_1  =  0$, and $c_2=1$. 
                      The regions where $g_\chi  >  1$ are shown as the gray region. 
                      The blue region is excluded by the latest result of the XENON1T experiment~\cite{1805.12562}.
                      The purple, brown, and gray dotted lines indicate the future sensitivities of the XENONnT~\cite{1512.07501}, LZ~\cite{1611.05525}, and DARWIN experiments~\cite{1606.07001}, respectively. 
                      The yellow region is below the neutrino floor~\cite{1307.5458}.
                      }
        \label{fig:mDM-range_demo}        
    \end{figure}

In Fig. \ref{fig:mDM-range_demo}, 
we show the predicted cross section in this model. 
We take $m_a  =  100$ GeV, $m_A  =  600$ GeV, $\theta = 0.1$, $t_\beta  =  10$, $c_1  =  0$, and $c_2=1$
and show the plot in $10~\text{GeV} <  m_\chi  <  1~\text{TeV}$. 
In the discontinuous part of the plot, 
we cannot find the solution of $g_\chi$ which realizes $\Omega  h^2  \sim  0.12$. 
The region where $g_\chi  >  1$ is shown as the gray region. 
The current bound, future prospects, and neutrino floor of the direct detection experiments are also shown in the figure. 
The blue regions are already excluded by the latest result of the XENON1T experiment~\cite{1805.12562}.
The purple, brown, and gray dotted lines indicate the future sensitivities of the XENONnT~\cite{1512.07501}, LZ~\cite{1611.05525}, and DARWIN experiments~\cite{1606.07001}, respectively. 
The yellow regions are below the neutrino floor~\cite{1307.5458}. 

As can be seen, 
we find that 
$g_\chi  > 1$ 
and the cross section is larger than the upper bound from the XENON1T experiment~\cite{1805.12562} 
in almost all of the  region where $m_\chi  <  (m_h  +  m_a)/2$ except for the funnel position of the light pseudoscalar, $m_\chi  \sim  m_a/2$. 
In addition, the previous work~\cite{1711.02110} has pointed out that the light DM mass region is excluded by the indirect detection experiments. 
From above reasons, 
we focus on the heavier DM mass region, $100$~GeV~$<~m_\chi~<~1$ TeV, 
and search the parameter space where the cross section becomes large. 

\begin{figure}[]
        \centering
        \begin{tabular}{c}
        \begin{minipage}{0.45\hsize}
            \centering
             \includegraphics[width=7cm]{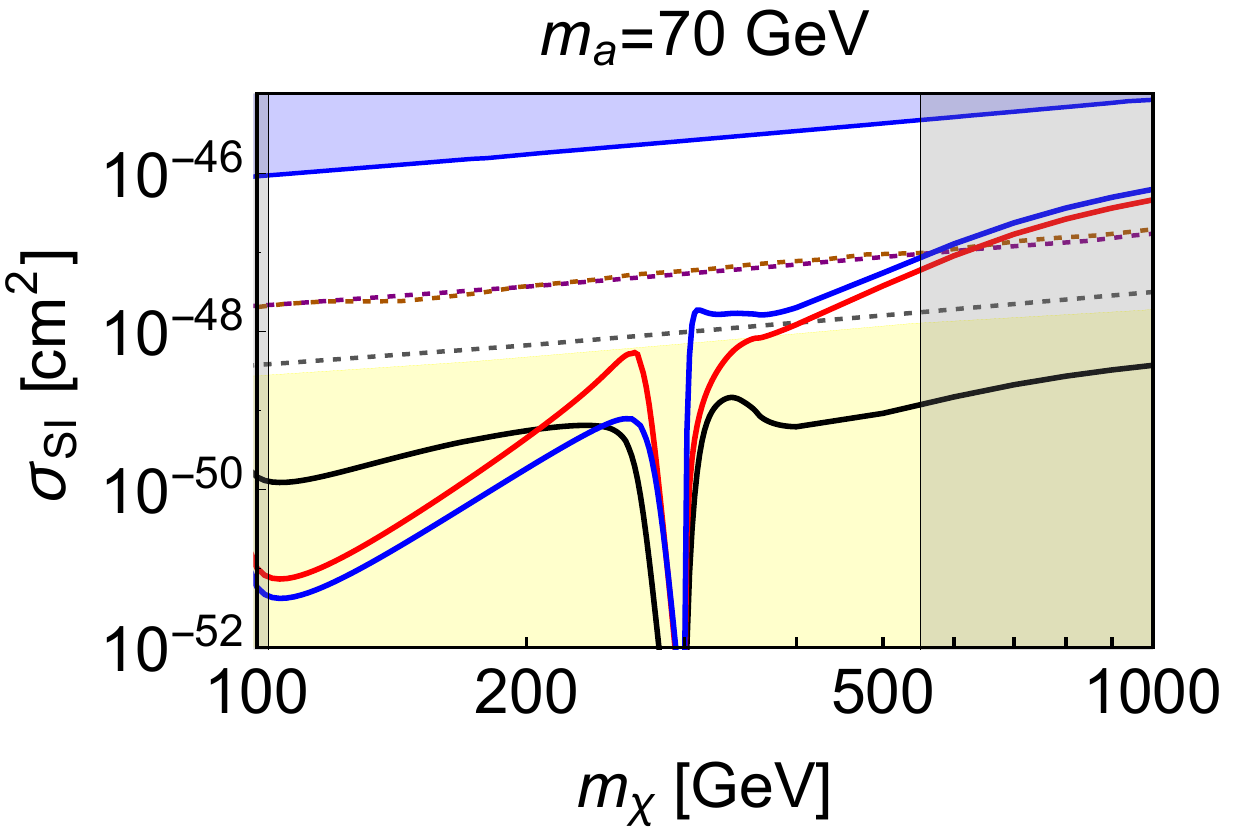}
        \end{minipage}
        \begin{minipage}{0.5\hsize}
            \centering
             \includegraphics[width=9.3cm]{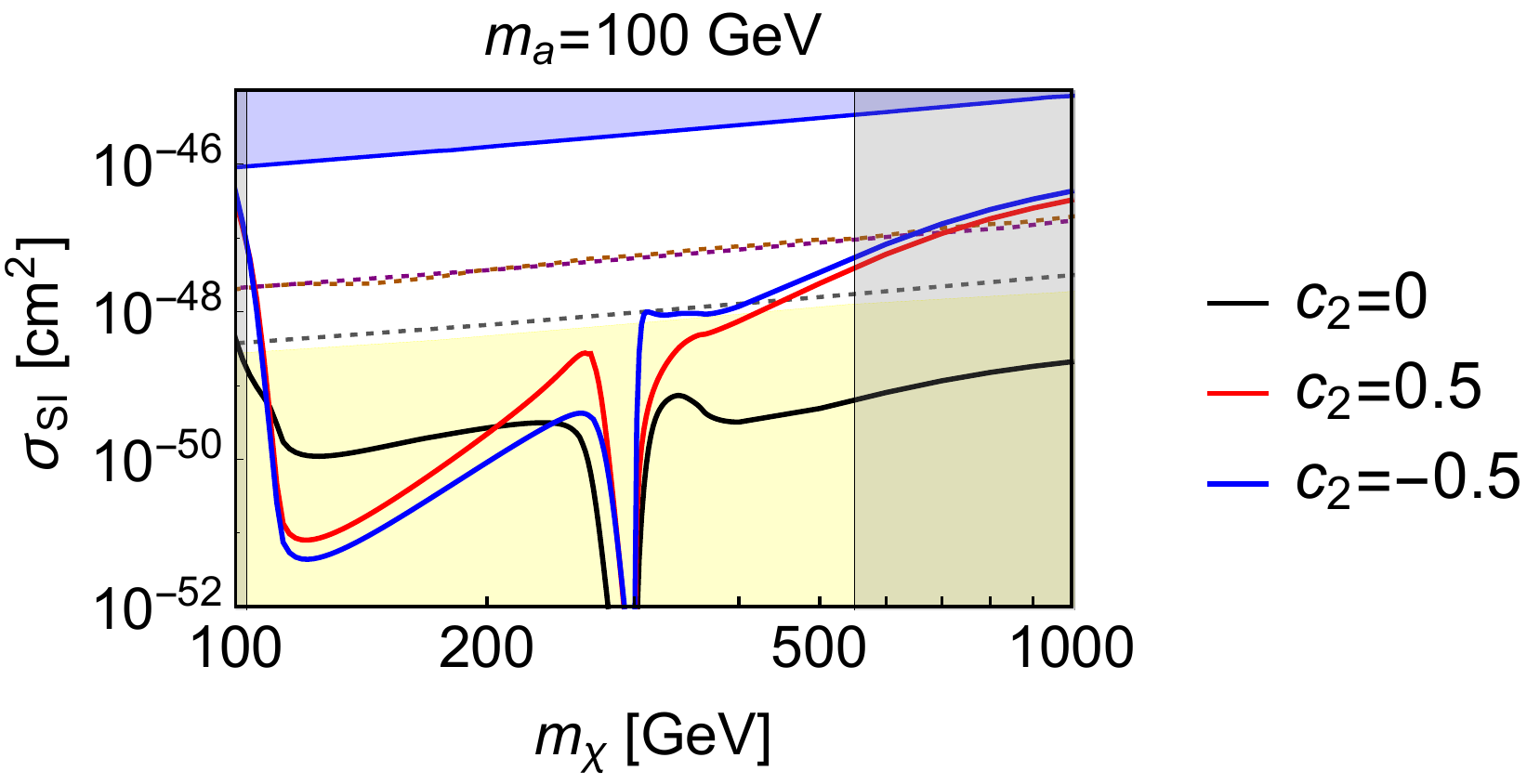}
        \end{minipage}
        \\
        \\
        \begin{minipage}{0.45\hsize}
            \centering
             \includegraphics[width=7cm]{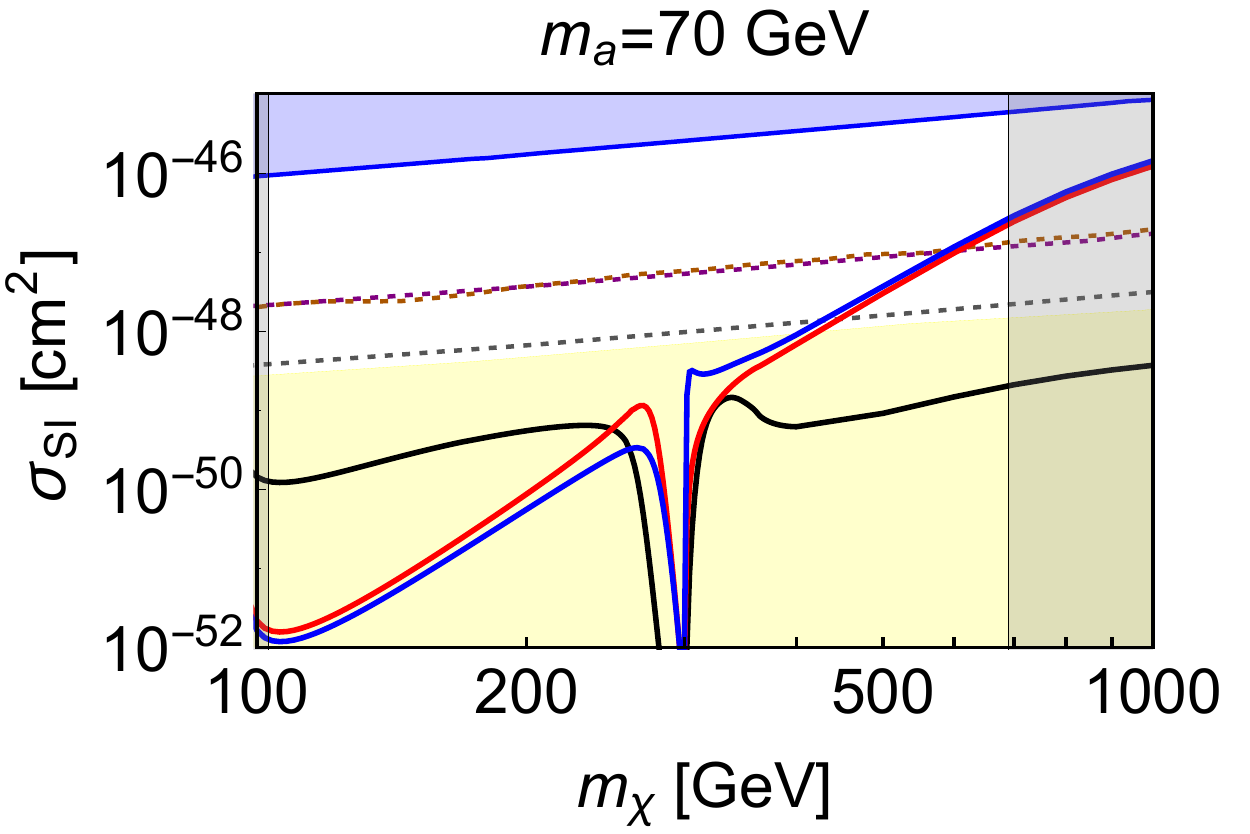}
        \end{minipage}
        \begin{minipage}{0.5\hsize}
            \centering
             \includegraphics[width=9cm]{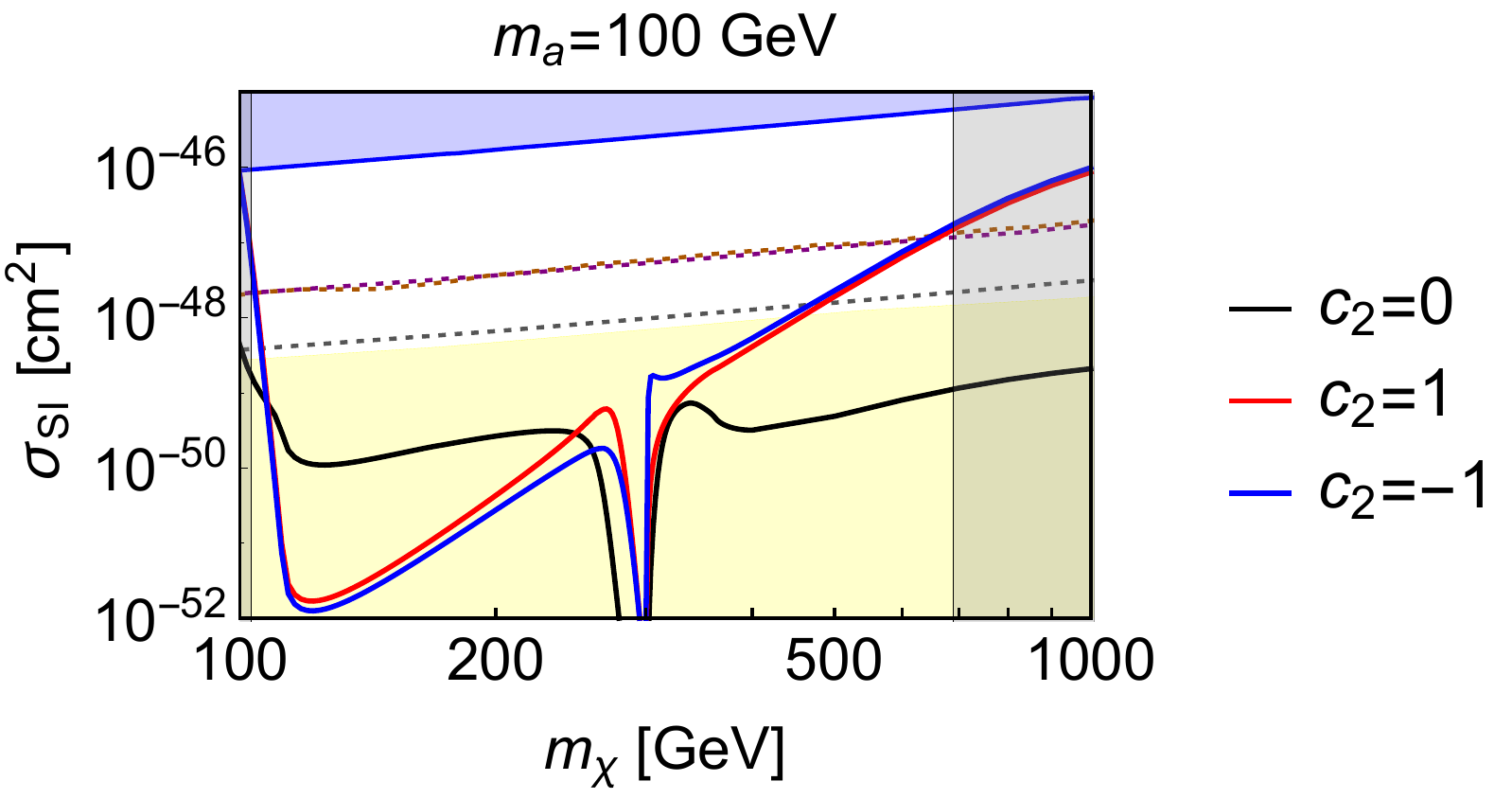}
        \end{minipage}
        \end{tabular}
        \caption{
                      The SI cross section at the four benchmark points. 
                      The upper panels show the case for $|c_2| = 0.5$, and the lower for $|c_2| = 1$. 
                      The left panels show the case for $m_a = 70$ GeV, and the right for $m_a = 100$ GeV. 
                      The other parameters are $m_A  =  600$ GeV, $\theta = 0.1$, $t_\beta  =  10$, $c_1  =  0$ for all the panels.
                      The regions where $g_\chi  >  1$ are shown as the gray region. 
                      The blue regions are excluded by the latest result of the XENON1T experiment~\cite{1805.12562}.
                      The purple, brown, and gray dotted lines indicate the future sensitivities of the XENONnT~\cite{1512.07501}, LZ~\cite{1611.05525}, and DARWIN experiments~\cite{1606.07001}, respectively. 
                      The yellow regions are below the neutrino floor~\cite{1307.5458}.
                      }
        \label{fig:comp_c2_ma}        
    \end{figure}

In Fig.~\ref{fig:comp_c2_ma}, we show the four benchmark points which have the different combinations of $c_2$ and $m_a$. 
Here, we fix the parameters at $m_A = 600$~GeV, $\theta=0.1$, $t_\beta =10$, and $c_1  =  0$. 
These benchmark points are all allowed from the current constraints as we mentioned in Sec. \ref{sec:model}.

We show the case for $|c_2|  =  0.5$ in the upper panels, and for $|c_2|  =  1$  in the lower panels. 
In each panel, the red line shows the positive $c_2$, the blue line shows the negative $c_2$, and the black line shows $c_2  =  0$.
We show the case for $m_a = 70$ GeV in the left panels, and $m_a = 100$~GeV in the right panels.
The DM coupling $g_\chi$ becomes larger than $1$ in the gray region. 
The current bound, future prospects, and background of the direct detection experiments shown in Fig.~\ref{fig:comp_c2_ma} are the same as those shown in Fig. \ref{fig:mDM-range_demo}.

From the figure, we find that $\sigma_{\rm SI}$ strongly depends on $c_2$. 
In particular, $\sigma_{\rm SI}$ becomes large for $m_\chi  \gtrsim  400$~GeV if $c_2$ is nonzero. 
This is because $g_{haa}$ becomes large by the effect of $c_2$ as we mentioned above. 
We also find that $\sigma_{\rm SI}$ for $m_a  =  70$~GeV is larger than that for $m_a  =  100$~GeV as is expected. 
At these benchmark points, 
we find the large region
where $\sigma_{\rm SI}$ is above the neutrino floor while keeping $g_\chi \leq 1$. 
For $c_2  =  1$, $m_a=70$ GeV, and $600$~GeV~$\leq~m_\chi~\leq~690$~GeV, 
$\sigma_{\rm SI}$ is above the future prospect lines of the XENONnT~\cite{1512.07501} and LZ experiments~\cite{1611.05525}
with $g_\chi \leq 1$.

We also have checked the cross section of the different types of the THDM
and found that $\sigma_{\rm SI}$ in the large $m_\chi$ region is type independent. 

\newpage
\section{Conclusions}
\label{sec:conclusion}
In this paper, 
we have discussed the physics of the DM direct detection in the pseudoscalar mediator DM model.
The tree-level amplitude of the DM-nucleon elastic scattering in this model is negligible 
because it is proportional to the momentum transfer in the non-relativistic limit. 
At the loop level, however, 
there are the diagrams which induce the DM-nucleon SI scattering. 
Thus, it is necessary to calculate the loop corrections to compare the model prediction with the direct detection experiments.

We have revisited the loop corrections to the cross section calculated in~\cite{1711.02110}
and improved their analysis with the following points.
For the triangle diagrams,
we read out the scalar trilinear couplings not only from $V_{\rm  port}$ but also from $V_{\rm  THDM}$ as shown in Eq.~\eqref{eq:trilinear}. 
We also included the diagrams with the heavy mediator $A$ into our analysis as pointed out in~\cite{1803.01574}. 
As a result, we found that the scattering amplitude of the triangle diagrams was overestimated in~\cite{1711.02110}. 
This is because the cancellation between the $haa$-diagram and the $haA$-diagram is not negligible numerically. 
For the box diagrams, 
we decomposed the effective operators into the scalar and twist-2 operators.
This decomposition gives the new contributions to the scalar operator, 
but we found that their effects are not significant.
In addition, we read out the DM-gluon scalar operator by calculating all the relevant two-loop diagrams.
We found that the contributions from the charm and bottom quarks to $C_G^{\rm box}$ were underestimated in~\cite{1711.02110}. 
On the other hand, the contribution from the top quark was overestimated. 
These results clearly show that it is no longer justified to relate $C_G^{\rm box}$ with $C_Q^{\rm box}$ using Eq.~\eqref{eq:qqtogg}.

In Sec.~\ref{sec:future}, we searched the region where the DM-nucleon scattering cross section becomes large. 
We found the two interesting cases. 
First, if $c_2$ is nonzero, the cross section is enhanced in the large $m_\chi$ region. 
This is because the contribution from $c_2$ to $g_{haa}$ appears without the suppression of the mixing angle and $t_\beta$. 
The interaction term proportional to $c_2$ is not included in the previous works,~\cite{1404.3716}, \cite{1711.02110} and~\cite{1803.01574}. 
Thus, our analysis has revealed the new possibility to detect the DM model with pseudoscalar mediators. 
Second, if $m_a$ is light, the cross section also becomes large because the suppression of the loop functions is weakened. 
In Fig.~\ref{fig:comp_c2_ma}, we showed the cross section at the four benchmark points. 
There are large regions where the cross section is above the neutrino floor while keeping the DM-pseudoscalar coupling perturbative.

The loop corrections in the scattering processes are often crucial.
The DM model of winos, the superpartner of SU(2)$_L$ gauge bosons, is one of the examples~\cite{Hisano:2004pv, 1004.4090, 1504.00915}.
In this model, the tree-level contribution is suppressed, 
and there are box diagrams which induce the SI scattering effects at the loop level. 
As pointed out in~\cite{1803.01574}, the same situation also happens in inelastic DM models 
which contain DM candidates with a tiny mass splitting. 
In these models, the two-loop calculations are necessary to evaluate the cross section 
and the same technique shown in this paper is available.

\section*{Acknowledgments}
This work was supported by JSPS KAKENHI Grant Number 16K17715 [T.A.]
and by Grant-in-Aid for Scientific research from the Ministry of
Education, Science, Sports, and Culture (MEXT), Japan, No. 16H06492
[J.H.]. The work of J.H. is also supported by World Premier
International Research Center Initiative (WPI Initiative), MEXT,
Japan.

\appendix  
\section{Scalar trilinear couplings}
\label{sec:3scalar}
The expressions of the scalar trilinear couplings in Eq.~\eqref{eq:trilinear} are as follows:
\begin{align}
    g_{haa}    =&    -  \frac{ 2  m_a^2 }{ v }  s_{ \beta  -  \alpha }  \sin^2 \theta  
                             -  \frac{ m_h^2 }{ v }  \sin^2 \theta  \frac{ s_{ \beta  -  \alpha }  t_\beta  +   c_{ \beta  -  \alpha }  ( 1  -  t_\beta^2 ) }{ t_\beta }
                             \nn   \\
                             &
                             +  \frac{ M^2 }{ v }  \sin^2 \theta  \frac{ 2  s_{ \beta  -  \alpha }  t_\beta  +  c_{ \beta  -  \alpha }  ( 1  -  t_\beta^2 ) }{ t_\beta }
                             \nn   \\
                             &
                             -2  c_1  v  \cos^2 \theta  \frac{ s_{ \beta  -  \alpha }  -  c_{ \beta  -  \alpha }  t_\beta }{ 1  +  t_\beta^2 }
                             -2  c_2  v  \cos^2 \theta  \frac{ t_\beta  ( s_{ \beta  -  \alpha }  t_\beta  +  c_{ \beta  -  \alpha }  ) }{ 1  +  t_\beta^2 }  
                             ,                             
                             \\
                             \nn   \\
    g_{haA}    =&    ~~ \frac{ m_A^2 }{ 2  v }  s_{ \beta  -  \alpha }  \sin  2  \theta
                             +  \frac{ m_a^2 }{ 2  v }  s_{ \beta  -  \alpha }   \sin  2  \theta
                             +  \frac{ m_h^2 }{ 2  v }  \sin  2  \theta  \frac{ s_{ \beta  -  \alpha }  t_\beta  +   c_{ \beta  -  \alpha }  ( 1  -  t_\beta^2 ) }{ t_\beta }
                             \nn   \\
                             &
                             -  \frac{ M^2 }{ 2  v }  \sin  2 \theta  \frac{ 2  s_{ \beta  -  \alpha }  t_\beta  +  c_{ \beta  -  \alpha }  ( 1  -  t_\beta^2 ) }{ t_\beta }
                             \nn   \\
                             &
                             -  c_1  v  \sin  2  \theta  \frac{ s_{ \beta  -  \alpha }  -  c_{ \beta  -  \alpha }  t_\beta }{ 1  +  t_\beta^2 }
                             -  c_2  v  \sin  2  \theta  \frac{ t_\beta  ( s_{ \beta  -  \alpha }  t_\beta  +  c_{ \beta  -  \alpha }  ) }{ 1  +  t_\beta^2 }  
                             ,                             
                             \\
                             \nn   \\
    g_{hAA}    =&    -  \frac{ 2  m_A^2 }{ v }  s_{ \beta  -  \alpha }  \cos^2 \theta  
                             -  \frac{ m_h^2 }{ v }  \cos^2 \theta  \frac{ s_{ \beta  -  \alpha }  t_\beta  +   c_{ \beta  -  \alpha }  ( 1  -  t_\beta^2 ) }{ t_\beta }
                             \nn   \\
                             &
                             +  \frac{ M^2 }{ v }  \cos^2 \theta  \frac{ 2  s_{ \beta  -  \alpha }  t_\beta  +  c_{ \beta  -  \alpha }  ( 1  -  t_\beta^2 ) }{ t_\beta }
                             \nn   \\
                             &
                             -2  c_1  v  \sin^2 \theta  \frac{ s_{ \beta  -  \alpha }  -  c_{ \beta  -  \alpha }  t_\beta }{ 1  +  t_\beta^2 }
                             -2  c_2  v  \sin^2 \theta  \frac{ t_\beta  ( s_{ \beta  -  \alpha }  t_\beta  +  c_{ \beta  -  \alpha }  ) }{ 1  +  t_\beta^2 }  
                             ,                    
                             \\
                             \nn   \\
    g_{Haa}    =&    -  \frac{ 2  m_a^2 }{ v }  c_{ \beta  -  \alpha }  \sin^2 \theta  
                             +  \frac{ m_H^2 }{ v }  \sin^2 \theta  \frac{ s_{ \beta  -  \alpha }  ( 1  -  t_\beta^2 )  -   c_{ \beta  -  \alpha }  t_\beta  }{ t_\beta }
                             \nn   \\
                             &
                             -  \frac{ M^2 }{ v }  \sin^2 \theta  \frac{ s_{ \beta  -  \alpha }  ( 1  -  t_\beta^2 )  -  2  c_{ \beta  -  \alpha }  t_\beta }{ t_\beta }
                             \nn   \\
                             &
                             -  2  c_1  v  \cos^2 \theta  \frac{ s_{ \beta  -  \alpha }  t_\beta  +  c_{ \beta  -  \alpha } }{ 1  +  t_\beta^2 }
                             +  2  c_2  v  \cos^2 \theta  \frac{ t_\beta  ( s_{ \beta  -  \alpha }  -  c_{ \beta  -  \alpha }  t_\beta  ) }{ 1  +  t_\beta^2 }  
                             ,                             
\end{align}
\begin{align}
    g_{HaA}    =&    \ \ \frac{ m_A^2 }{ 2  v }  c_{ \beta  -  \alpha }  \sin  2  \theta
                             +  \frac{ m_a^2 }{ 2  v }  c_{ \beta  -  \alpha }   \sin  2  \theta
                             -  \frac{ m_H^2 }{ 2  v }  \sin  2  \theta  \frac{ s_{ \beta  -  \alpha }  ( 1  -  t_\beta^2 )  -  c_{ \beta  -  \alpha }  t_\beta }{ t_\beta }
                             \nn   \\
                             &
                             +  \frac{ M^2 }{ 2  v }  \sin  2  \theta  \frac{ s_{ \beta  -  \alpha }  ( 1  -  t_\beta^2 )  -  2  c_{ \beta  -  \alpha }  t_\beta }{ t_\beta }
                             \nn   \\
                             &
                             -  c_1  v  \sin  2  \theta  \frac{ s_{ \beta  -  \alpha }  t_\beta  +  c_{ \beta  -  \alpha } }{ 1  +  t_\beta^2 }
                             +  c_2  v  \sin  2  \theta  \frac{ t_\beta  ( s_{ \beta  -  \alpha }  -  c_{ \beta  -  \alpha }  t_\beta ) }{ 1  +  t_\beta^2 }  
                             ,                             
                             \\
                             \nn   \\
    g_{HAA}    =&    -  \frac{ 2  m_A^2 }{ v }  c_{ \beta  -  \alpha }  \cos^2 \theta  
                             +  \frac{ m_H^2 }{ v }  \cos^2 \theta  \frac{ s_{ \beta  -  \alpha }  ( 1  -  t_\beta^2 )  +   c_{ \beta  -  \alpha }  t_\beta }{ t_\beta }
                             \nn   \\
                             &
                             -  \frac{ M^2 }{ v }  \cos^2 \theta  \frac{ s_{ \beta  -  \alpha }  ( 1  -  t_\beta^2 )  -  2  c_{ \beta  -  \alpha }  t_\beta }{ t_\beta }
                             \nn   \\
                             &
                             -2  c_1  v  \sin^2 \theta  \frac{ s_{ \beta  -  \alpha }  t_\beta  -  c_{ \beta  -  \alpha } }{ 1  +  t_\beta^2 }
                             +2  c_2  v  \sin^2 \theta  \frac{ t_\beta  ( s_{ \beta  -  \alpha }  -  c_{ \beta  -  \alpha }  t_\beta ) }{ 1  +  t_\beta^2 },
\end{align}
where
\begin{align}
M^2    =    \frac{ m^2_{ 3 } }{ \sin \beta  \cos \beta }, \quad
s_{ \beta  -  \alpha }=\sin( \beta  -  \alpha ),  \quad
c_{ \beta  -  \alpha }=\cos( \beta  -  \alpha ). 
\end{align} 
Taking the alignment limit, $\sin(\beta - \alpha) \to 1$, we find
\begin{align}
    g_{haa}    =&    \ \ \ (  2  M^2  -  2  m_a^2  -  m_h^2 )  \frac{ \sin^2  \theta }{ v }
                             -  2  v  \frac{ c_1  +  c_2  t_\beta^2 }{ 1  +  t_\beta^2 }  \cos^2 \theta
                             ,                             
                             \label{eq:ghaa}
                             \\
                             \nn   \\
    g_{haA}    =&    -  (  2  M^2  -  m_a^2  -  m_A^2  -  m_h^2 )  \frac{ \sin  2  \theta }{ 2  v }    
                             -  v  \frac{ c_1  +  c_2  t_\beta^2 }{ 1  +  t_\beta^2 }  \sin  2  \theta
                             ,                             
                             \\
                             \nn   \\
    g_{hAA}    =&    \ \ \ (  2  M^2  -  2  m_A^2  -  m_h^2 )  \frac{ \cos^2  \theta }{ v }
                             -  2  v  \frac{ c_1  +  c_2  t_\beta^2 }{ 1  +  t_\beta^2 }  \sin^2 \theta
                             ,                             
                             \\
                             \nn   \\
    g_{Haa}    =&    \ \ \ ( M^2  -  m_H^2 )  \frac{ \sin^2 \theta }{ v }\Bigl( 1  -  \frac{ 1 }{ t_\beta } \Bigr)
                             -  2  v  ( c_1  -  c_2 )  \frac{ t_\beta }{ 1  +  t_\beta^2 }  \cos^2 \theta
                             ,                             
                             \\
                             \nn   \\
    g_{HaA}    =&   ( m_H^2  -  M^2 )  \frac{ \sin 2 \theta }{ 2  v }  \Bigl( 1  -  \frac{ 1 }{ t_\beta } \Bigr)
                             - v  ( c_1  -  c_2 )  \frac{ t_\beta }{ 1  +  t_\beta^2 }  \sin 2 \theta
                             ,                             
                             \\
                             \nn   \\
    g_{HAA}    =&   \ \ \ ( M^2  -  m_H^2 )  \frac{ \cos^2  \theta }{ v }  \Bigl( 1  -  \frac{ 1 }{ t_\beta } \Bigr)
                             -  2  v  ( c_1  -  c_2 )  \frac{ t_\beta }{ 1  +  t_\beta^2 }    \sin^2 \theta
                             .
\end{align}
In the previous work~\cite{1711.02110}, they set $m_H^2  =  m_A^2  =  M^2$ and $c_1  =  c_2  =  0$. 
In this case, we find $g_{Ha_i a_j} =  0$ where $a_i = a, A$.

\section{Details of the calculations of the box diagrams}
\label{sec:boxdetail}
In the following, we show how to derive the effective operators from the box diagrams. 
\subsection{The derivation of $C_q^{\rm box}$, $C_q^{(1)\rm box}$, and $C_q^{(2)\rm box}$}
\label{sec:box_cq_detail}
        \begin{figure}[]
          \begin{center}
            \begin{tabular}{c}
              \begin{minipage}{0.4\hsize}
                \begin{center}
                \includegraphics[width=5cm]{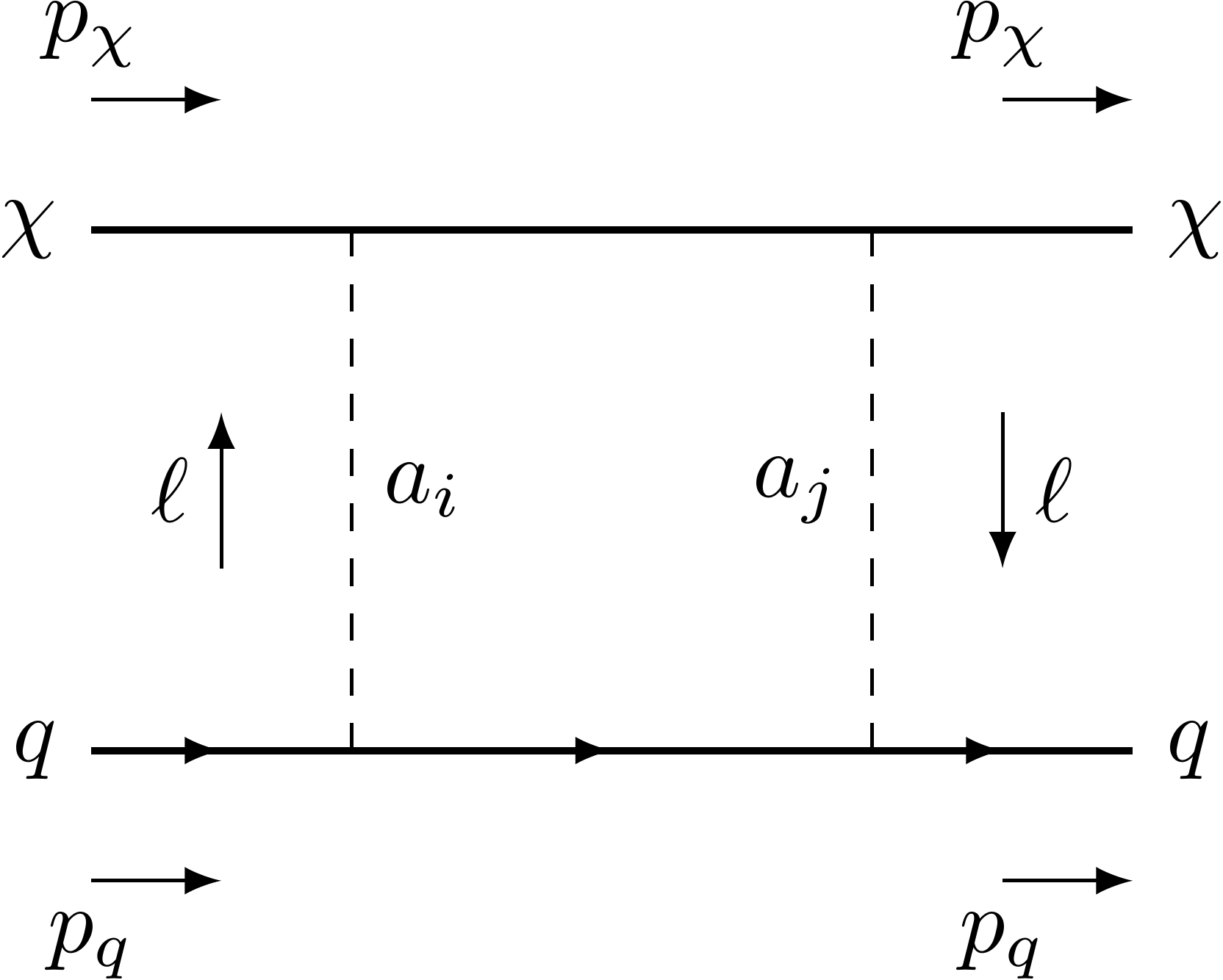}
                \end{center}
              \end{minipage}
              \begin{minipage}{0.4\hsize}
                \begin{center}
                \includegraphics[width=5cm]{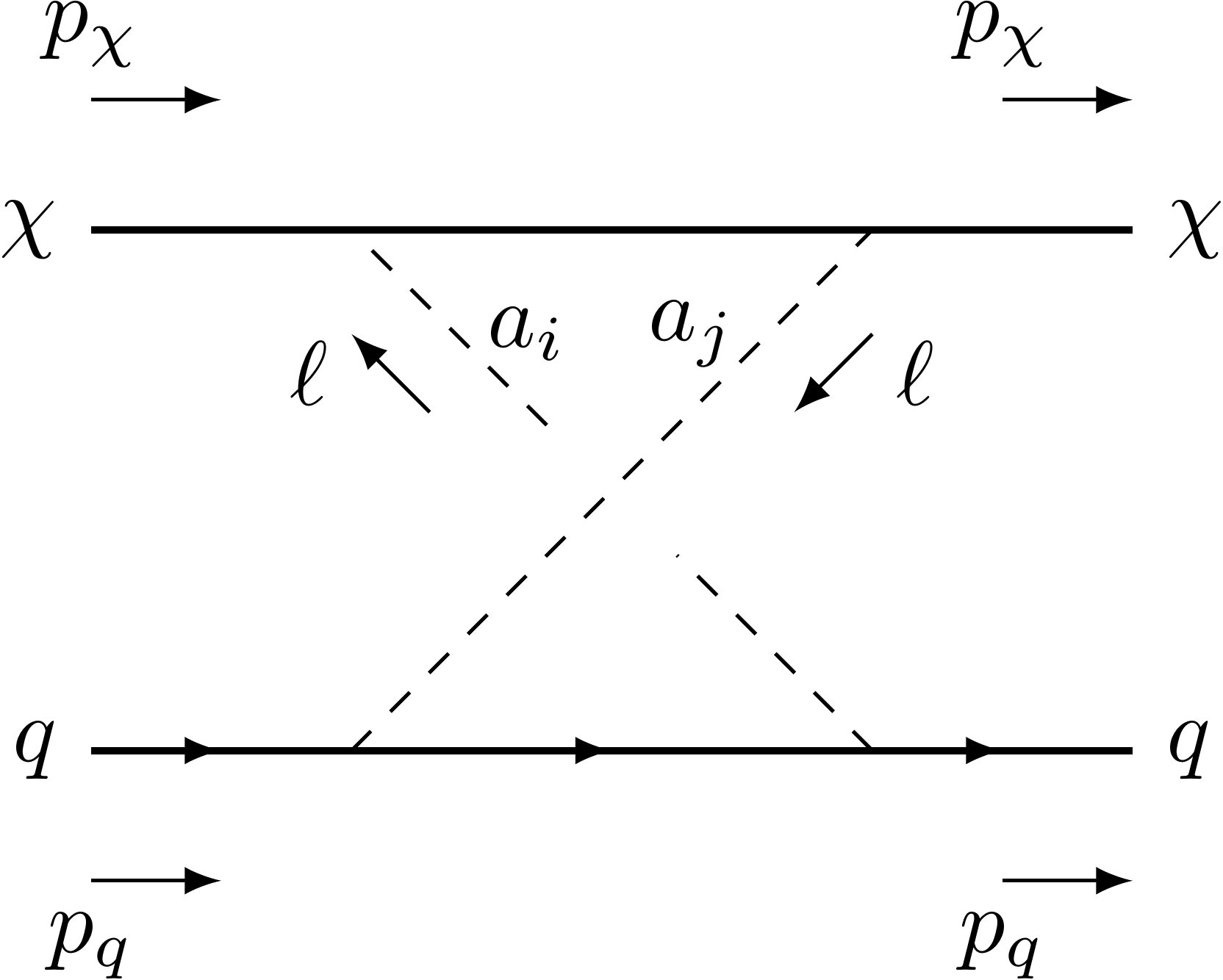}
                \end{center}
              \end{minipage}
            \end{tabular}
          \end{center}
          \caption{The box diagrams and their momentum assignments. }
          \label{fig:boxmomentum}
        \end{figure}
We show the details of the calculations of $C_q^{\rm box}$, $C_q^{(1)\rm box}$, and $C_q^{(2)\rm box}$ shown in Sec.~\ref{sec:boxlq}. 
Note that we calculate the amplitude of the DM-quark scattering process in the zero momentum transfer limit. 
Summing up the amplitude of the box diagrams shown in Fig.~\ref{fig:boxmomentum}, we obtain
\begin{align}
i  {\cal  M}  =
\left( \frac{ m_q }{ v } \right)^2  \xi^\chi_{a_i}  \xi^\chi_{a_j}  \xi^q_{a_i}  \xi^q_{a_j}
\Bigl[ \bar{u}_\chi(p_\chi) \gamma^\mu u_\chi(p_\chi) \Bigr]  \Bigl[ \bar{u}_q(p_q) \gamma^\nu u_q(p_q) \Bigr]&
\nn\\
\times
\int\frac{ d^D \ell }{ (2\pi)^D }
\frac{ \ell_\mu \ell_\nu}{\left[(\ell+p_\chi)^2 - m_\chi^2\right]  ( \ell^2 - m_{a_i}^2 )  ( \ell^2 - m_{a_j}^2 ) }&
\nn   \\
\times
\Bigl[ \frac{-1}{(\ell-p_q)^2 - m_q^2}  +  &\frac{1}{ (\ell  +  p_q)^2 - m_q^2 } \Bigr]
,
\label{eq:amp_box_1}
\end{align}
where $u_\chi(p_\chi)$ is the DM wave function with its momentum $p_\chi$, 
and $u_q(p_q)$ is the quark wave function with its momentum $p_q$. 
We expand the terms in the bracket in Eq.~\eqref{eq:amp_box_1} by the quark momentum and keep the leading term as follows: 
\begin{align}
\frac{-1}{(\ell-p_q)^2 - m_q^2}  +  \frac{1}{ (\ell  +  p_q)^2 - m_q^2 }
    &=
    -  \frac{4  \ell  \cdot  p_q}{\ell^4}  +  {\cal  O}(p_q^2).
\end{align}
Here we have used $p_q^2  =  m_q^2$. 
After this expansion, the amplitude of these diagrams is  
\begin{align}
    i  { \cal  M }    =&~~~~    i  A_1    m_\chi  m_q  \Bigl[ \bar{ u }_\chi ( p_\chi )  u_\chi ( p_\chi ) \Bigr]  \Bigl[ \bar{ u }_q ( p_q )  u_q ( p_q ) \Bigr]  \nn   \\
                            &    +     i  A_1    ( p_\chi  \cdot  p_q )  \Bigl[ \bar{ u }_\chi ( p_\chi )  \gamma^\mu  u_\chi ( p_\chi ) \Bigr]  \Bigl[ \bar{ u }_q ( p_q )  \gamma_\mu  u_q ( p_q ) \Bigr]  \nn   \\
                            &    +     i  A_1    \Bigl[ \bar{ u }_\chi ( p_\chi )  \slashed{p}_q  u_\chi ( p_\chi ) \Bigr]  \Bigl[ \bar{ u }_q ( p_q )  \slashed{p}_\chi  u_q ( p_q ) \Bigr]  \nn   \\
                            &    +     i  A_2    m_\chi  ( p_\chi  \cdot  p_q )  \Bigl[ \bar{ u }_\chi ( p_\chi )  u_\chi ( p_\chi ) \Bigr]  \Bigl[ \bar{ u }_q ( p_q )  \slashed{p}_\chi  u_q ( p_q ) \Bigr],
                            \label{eq:amp_box}
\end{align}
where
\begin{align}
    A_1  =    \frac{ -  4 }{ ( 4  \pi )^2 }  \left( \frac{ m_q }{ v } \right)^2
                    \times  
                    \Biggl\{ 
                        &    \frac{ ( \xi^\chi_a  \xi^q_a )^2 }{ m_a^2 }  \left[ X_{001} ( p^2,  m_\chi^2,  0,  m_a^2 )  -  X_{001} ( p^2,  m_\chi^2,  m_a^2,  0 ) \right]
                        \nn   \\
                        &    +
                        \frac{ ( \xi^\chi_A  \xi^q_A )^2 }{ m_A^2 }  \left[ X_{001} ( p^2,  m_\chi^2,  0,  m_A^2 )  -  X_{001} ( p^2,  m_\chi^2,  m_A^2,  0 ) \right]
                        \nn   \\
                        &    +
                        2  \frac{ \xi^\chi_A  \xi^\chi_a  \xi^q_A  \xi^q_a }{ m_A^2  -  m_a^2 }  \left[ X_{001} ( p^2,  m_\chi^2,  m_A^2,  0 )  -  X_{001} ( p^2,  m_\chi^2,  m_a^2,  0 ) \right]
                    \Biggr\},
                    \label{eq:a1box}
                    \\
                    \nn   \\
    A_2  =    \frac{ -  4 }{ ( 4  \pi )^2 }  \left( \frac{ m_q }{ v } \right)^2
                    \times  
                    \Biggl\{ 
                        &    \frac{ ( \xi^\chi_a  \xi^q_a )^2 }{ m_a^2 }  \left[ X_{111} ( p^2,  m_\chi^2,  0,  m_a^2 )  -  X_{111} ( p^2,  m_\chi^2,  m_a^2,  0 ) \right]
                        \nn   \\
                        &    +
                        \frac{ ( \xi^\chi_A  \xi^q_A )^2 }{ m_A^2 }  \left[ X_{111} ( p^2,  m_\chi^2,  0,  m_A^2 )  -  X_{111} ( p^2,  m_\chi^2,  m_A^2,  0 ) \right]
                        \nn   \\
                        &    +
                        2  \frac{ \xi^\chi_A  \xi^\chi_a  \xi^q_A  \xi^q_a }{ m_A^2  -  m_a^2 }  \left[ X_{111} ( p^2,  m_\chi^2,  m_A^2,  0 )  -  X_{111} ( p^2,  m_\chi^2,  m_a^2,  0 ) \right]
                    \Biggr\}. 
                    \label{eq:a4box}
\end{align}
Loop functions $X_{001}$ and $X_{111}$ are defined in Appendix \ref{sec:loopfun_x}. 
From this amplitude, we find the following effective operators. 
\begin{align}
    { \cal  L }_{ \rm  eff }    =&          \frac{ 1 }{ 2 }     ( m_\chi  A_1 )  m_q  \bar{ \chi }      \chi  \bar{ q }  q
                                                +    \frac{ 1 }{ 2 }     A_1      ( \bar{ \chi }  i  \del^\mu  \gamma^\nu  \chi )  ( \bar{ q }  i  \del_\mu  \gamma_\nu  q )
                                                +    \frac{ 1 }{ 2 }     A_1      ( \bar{ \chi }  i  \del^\mu  \gamma^\nu  \chi )  ( \bar{ q }  i  \del_\nu  \gamma_\mu  q )
                                                \nn   \\
                                                &
                                                +    \frac{ 1 }{ 2 }    ( m_\chi  A_2 )  ( \bar{ \chi }  i  \del^\mu  i  \del^\nu  \chi )  ( \bar{ q }  i  \del_\mu  \gamma_\nu  q ). 
\end{align}
Then, we perform the irreducible decomposition to these operators. 
\begin{align}
              \bar{ q }  i  \del^\mu  \gamma^\nu  q    
    &=     \bar{ q }     \left[ \frac{ i  \del^\mu  \gamma^\nu  +  i  \del^\nu  \gamma^\mu }{ 2 }  -  \frac{1}{4}  g^{\mu \nu}  i  \slashed{\del} \right]  q
              +
              \bar{ q }    \left[ \frac{ i  \del^\mu  \gamma^\nu  -  i  \del^\nu  \gamma^\mu }{ 2 }  \right]  q
              +
              \frac{ 1 }{ 4 }  g^{ \mu  \nu }  \bar{ q }  i  \slashed{\del}  q
 \nn\\
    &=     { \cal  O }_{ \mu  \nu }^q
              +
              \frac{1}{4} g^{\mu \nu} m_q \bar{ q }  q .
              \label{eq:irreducible}
\end{align}
Note that we drop the anti-symmetric term in the last line 
because it does not contribute to the nucleon matrix element. 
The last term in Eq.~(\ref{eq:irreducible}) gives the contribution to the scalar operator $\bar{\chi}  \chi  \bar{ q }  q$. 
After this decomposition, we find
\begin{align}
   { \cal  L }_{ \rm  eff }    =&          \frac{ 1 }{ 2 }     C_q^{ \rm box }                 m_q  \bar{ \chi }      \chi  \bar{ q }  q
                                                +    \frac{ 1 }{ 2 }    C_q^{ ( 1 )  { \rm box } }    \bar{ \chi }  i  \del^\mu  \gamma^\nu  \chi  { \cal  O }_{ \mu  \nu }^q
                                                +    \frac{ 1 }{ 2 }    C_q^{ ( 2 )  { \rm box } }    \bar{ \chi }  i  \del^\mu  i  \del^\nu  \chi       { \cal  O }_{ \mu  \nu }^q,
\end{align}
where
\begin{align}
    C_q^{ \rm box }                 =&    \frac{m_\chi}{4}  \left(  6  A_1  +  m_\chi^2  A_2   \right),  
                                              \\
    C_q^{ ( 1 )  { \rm box } }     =&    2  A_1,
                                               \\
    C_q^{ ( 2 )  { \rm box } }     =&    m_\chi  A_2.
\end{align}
These coefficients correspond to the Wilson coefficients in Eqs.~(\ref{eq:scalarwc})--(\ref{eq:twist2wc}).

\subsection{The derivation of $C_G^{\rm box}$}
\label{sec:box_2loop_detail}
        \begin{figure}[]
          \begin{center}
                \includegraphics[width=14cm]{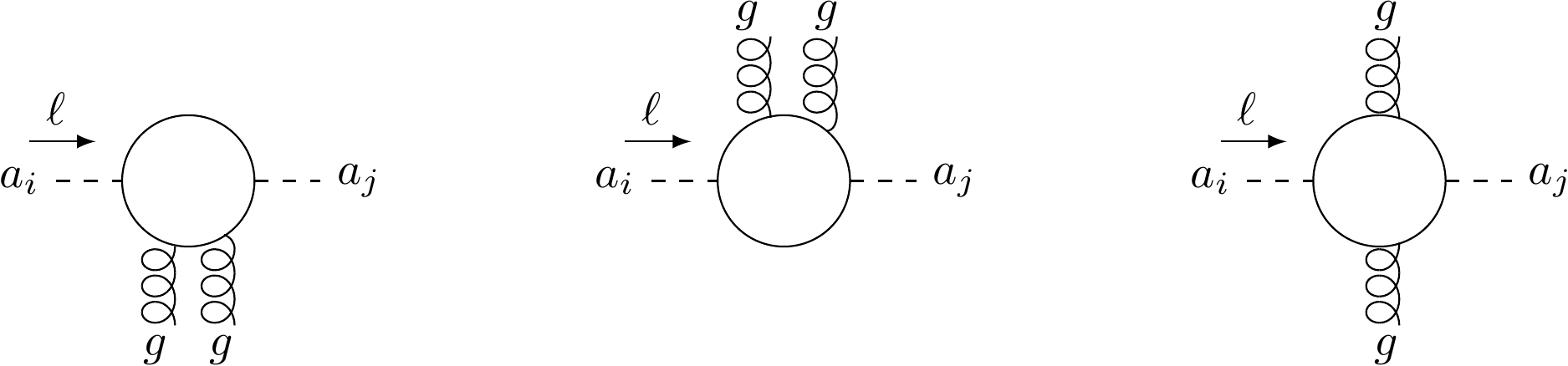}
          \end{center}
          \caption{The diagrams which contribute to the effective operators $a_i  a_j  G^a_{ \mu \nu }  G^{ a \mu \nu }$, where $a_i  =  a, A$.}
          \label{fig:aagg_diagram}
        \end{figure}
We show the details of the derivation of $C_G^{\rm box}$ shown in Sec.~\ref{sec:box_glue}.

First, we calculate the heavy quark loops in the two-loop diagrams. 
In Fig.~\ref{fig:aagg_diagram}, we show the pseudoscalar-gluon scattering diagrams. 
We calculate the amplitude of these diagrams in the gluon background field theory.  
We use the Fock-Schwinger gauge for the gluon field. 
The amplitude with the external pseudoscalar momentum $\ell$ is as follows: 
\begin{align}
    i  {\cal  M}   
    =&           
    i  \Pi_{a_i  a_j}  ( \ell^2 )  \Bigl[ G_{ \mu  \nu }^a  G^{ a  \mu  \nu } \Bigr],
    \label{eq:amp_aaGG}
\end{align}
where
\begin{align}
     \Pi_{a_i  a_j}  ( \ell^2 )
    &=    \frac{ \alpha_s }{ 24  \pi }  \Bigl[ G_{ \mu  \nu }^a  G^{ a  \mu  \nu } \Bigr]
             \sum_{Q=c, b, t}\left( \frac{ m_Q }{ v } \right)^2  \xi^Q_{a_i}  \xi^Q_{a_j}
             \nn   \\
        &  ~~\times
             \int_0^1 \! \! dx 
             \left[
               \scalebox{0.93}{$\displaystyle \frac{  3  x  ( 1  -  x ) }{ [ m_Q^2  -  x  ( 1  -  x )  \ell^2 ] } $}
               +
               \scalebox{0.93}{$\displaystyle \frac{  m_Q^2  \cdot  ( 2  +  5  x  -  5  x^2 ) }{ [ m_Q^2  -  x  ( 1  -  x )  \ell^2 ]^2 } $}
               -
               \scalebox{0.93}{$\displaystyle \frac{ 2  m_Q^4  ( 1  -  2  x  +  2  x^2 ) }{ [ m_Q^2  -  x  ( 1  -  x )  \ell^2 ]^3 } $}
             \right]. 
\end{align}
Note that the gluon in Eq.~\eqref{eq:amp_aaGG} is the external field.

Next, we read out the effective operator $a_i  a_j  G^a_{ \mu \nu }  G^{ a \mu \nu }$ from Eq.~\eqref{eq:amp_aaGG}
and calculate the amplitude of the two-loop diagrams shown in Fig.~\ref{fig:triboxcomp}. 
The amplitude can be expressed using $\Pi_{a_i  a_j}  ( \ell^2 )$ as follows: 
\begin{align}
    i  {\cal  M}   
    &=   
    \sum_{a_i = a, A}
     ( -  \xi^{\chi}_{a_i}  \xi^{\chi}_{a_j} )
     \Bigl[ \ol{u}_\chi  ( p_\chi )  \gamma^\rho  u_\chi( p_\chi )  \Bigr]  \Bigl[ G_{ \mu  \nu }^a  G^{ a  \mu  \nu } \Bigr]  \nn   \\
     &~~~~~~
     \times
        \int  \frac{ d^D \ell }{ (2  \pi)^D }  \  
        \frac{ \ell_\rho }{ [ ( \ell  +  p_\chi )^2  -  m_{ \chi }^2 ]  ( \ell^2  -  m_{a_i}^2)  ( \ell^2  -  m_{a_j}^2 ) }  
        \Pi_{a_i  a_j} ( \ell^2 ).
\end{align}
Reading out the effective operator from this amplitude, we obtain $C_G^{\rm box}$ in Eq.~\eqref{eq:cg_box}.

\section{Loop functions}
\label{sec:appendix_loop}
In the following, we show the expressions of the loop functions used in Sec.~\ref{sec:eo}. 
For the later convenience, we define $\tilde{\epsilon}$ as follows:  
    \begin{align}
        \frac{2}{\tilde{\epsilon}}  =  \frac{2}{\epsilon}  -  \gamma_E  +  \log (4\pi), 
    \end{align}
    where $\epsilon=4-D$ and $\gamma_E$ is the Euler-Mascheroni constant.
    \subsection{$B$ functions}
    \label{sec:loopfun_b}
    We show the definitions of $B$ functions ($B_0, B_1, B_{00}, B_{11}, B_{001}, B_{111}$) which are the same as those in \texttt{LoopTools}~\cite{Hahn:1998yk}. 
    \begin{align}
        &\int  \frac{ d^D \ell }{ (2  \pi)^D }  \   \frac{ 1                          }{ ( \ell^2  -  m^2 )  \left[  ( \ell  +  p )^2  -  M^2  \right] }        =        \frac{ i }{ ( 4  \pi )^2 }  B_0  ( p^2,  m^2,  M^2 ),    
        \\
        \nn   \\
        &\int  \frac{ d^D \ell }{ (2  \pi)^D }  \   \frac{ \ell_\mu               }{ ( \ell^2  -  m^2 )  \left[  ( \ell  +  p )^2  -  M^2  \right] }         =        \frac{ i }{ ( 4  \pi )^2 }  p_\mu  B_1  ( p^2,  m^2,  M^2 ),     
        \\
        \nn   \\
        &\int  \frac{ d^D \ell }{ (2  \pi)^D }  \   \frac{ \ell_\mu  \ell_\nu }{ ( \ell^2  -  m^2 )  \left[  ( \ell  +  p )^2  -  M^2  \right] }  
        \nn   \\       
        &~~~~~~=        \frac{ i }{ ( 4  \pi )^2 }  \left[  g_{ \mu  \nu }  B_{ 00 }  ( p^2,  m^2,  M^2 )  +  p_\mu  p_\nu  B_{ 11 }  ( p^2,  m^2,  M^2 )  \right],      
        \\
        \nn   \\
        &\int  \frac{ d^D \ell }{ (2  \pi)^D }  \   \frac{ \ell_\mu  \ell_\nu  \ell_\rho }{ ( \ell^2  -  m^2 )  \left[ ( \ell  +  p )^2  -  M^2 \right] } 
        \nn   \\
        &~~~~~~=    \frac{ i }{ ( 4  \pi )^2 }  \biggl[ ( g_{ \mu  \nu }  p_\rho  +  g_{ \nu  \rho }  p_\mu  +  g_{ \rho  \mu }  p_\nu )  B_{ 001 }  ( p^2,  m^2,  M^2 )  +  p_\mu  p_\nu  p_\rho  B_{ 111 }  ( p^2,  m^2,  M^2 ) \biggl].
    \end{align}
    The expressions of the $B$ functions are
    \begin{align}
        B_0  ( p^2,  m^2,  M^2 )            &=        \int_0^1  dx  \left[ \frac{ 2 }{ \tilde{\epsilon} }  +  \log  \left(  \frac{ \mu^2 }{ \scalebox{0.9}{$\displaystyle m^2  x  +  M^2  ( 1  -  x )  -  p^2  x  ( 1  -  x ) $} }   \right)  \right],        
                                                                      \\
                                                                      \nn   \\
        B_1  ( p^2,  m^2,  M^2 )            &=        \int_0^1  dx  [ -  ( 1  -  x ) ]  \left[ \frac{ 2 }{ \tilde{\epsilon} }  +  \log  \left(  \frac{ \mu^2 }{ \scalebox{0.9}{$\displaystyle m^2  x  +  M^2  ( 1  -  x )  -  p^2  x  ( 1  -  x ) $} }  \right)  \right],
                                                                      \\
                                                                      \nn   \\
        B_{00}  ( p^2,  m^2,  M^2 )        &=        \int_0^1  dx  \frac{ \scalebox{0.9}{$\displaystyle m^2  x  +  M^2  ( 1  -  x )  -  p^2  x  ( 1  -  x ) $} }{ 2 }  
                                                                      \nn   \\
                                                                      &  ~~~~~~~~~~~~~~\times
                                                                      \left[ \frac{ 2 }{ \tilde{\epsilon} }  +  1  +  \log  \left( \frac{ \mu^2 }{ \scalebox{0.9}{$\displaystyle m^2  x  +  M^2  ( 1  -  x )  -  p^2  x  ( 1  -  x ) $} }  \right)  \right],
                                                                      \\
                                                                      \nn   \\
        B_{11}  ( p^2,  m^2,  M^2 )        &=        \int_0^1  dx  ( 1  -  x )^2  \left[ \frac{ 2 }{ \tilde{\epsilon} }  +  \log  \left(  \frac{ \mu^2 }{ \scalebox{0.9}{$\displaystyle m^2  x  +  M^2  ( 1  -  x )  -  p^2  x  ( 1  -  x ) $} }  \right)  \right],
                                                                      \\
                                                                      \nn   \\
        B_{001}  ( p^2,  m^2,  M^2 )      &=        \int_0^1  dx  \frac{ -  ( 1  -  x )  [ \scalebox{0.9}{$\displaystyle m^2  x  +  M^2  ( 1  -  x )  -  p^2  x  ( 1  -  x ) $} ] }{ 2 }  
                                                                       \nn   \\
                                                                       &~~~~~~~~~~~~~~
                                                                       \times
                                                                       \left[ \frac{ 2 }{ \tilde{\epsilon} }  +  1  +  \log  \left(  \frac{ \mu^2 }{ \scalebox{0.9}{$\displaystyle m^2  x  +  M^2  ( 1  -  x )  -  p^2  x  ( 1  -  x ) $} }  \right)  \right],
                                                                      \\
                                                                      \nn   \\  
        B_{111}  ( p^2,  m^2,  M^2 )       &=        \int_0^1  dx  [ -  ( 1  -  x )^3 ]  \left[ \frac{ 2 }{ \tilde{\epsilon} }  +  \log  \left(  \frac{ \mu^2 }{ \scalebox{0.9}{$\displaystyle m^2  x  +  M^2  ( 1  -  x )  -  p^2  x  ( 1  -  x ) $} }  \right)  \right].
    \end{align}
    The derivative of $B_0$ with respect to $p^2$ are
    \begin{align}
        \frac{ \del }{ \del  p^2 }  B_0  ( p^2,  m^2,  M^2 )            
                                                          &=        \int_0^1  dx \frac{ x  ( 1  -  x ) }{ \scalebox{0.9}{$\displaystyle m^2  x  +  M^2  ( 1  -  x )  -  p^2  x  ( 1  -  x ) $} }.
    \end{align}

    \subsection{$X$ functions}
    \label{sec:loopfun_x}
    The definitions of $X$ functions ($X_{001}, X_{111}$) are as follows:
    \begin{align}
                 \int  &\frac{ d^D \ell }{ (2  \pi)^D }  \frac{ \ell_\mu  \ell_\nu  \ell_\rho }{\left[( \ell  +  p )^2  -  M^2 \right]  ( \ell^2  -  m_1^2 )^2  ( \ell^2  -  m_2^2 ) }  \nn   \\
        &=    \frac{ i }{ ( 4  \pi )^2 }  
                 \Bigl[ 
                 \scalebox{0.94}{$\displaystyle
                   ( g_{ \mu  \nu }  p_\rho  +  g_{ \nu  \rho }  p_\mu  +  g_{ \rho  \mu }  p_\nu )  X_{ 001 }  ( p^2,  M^2,  m_1^2,  m_2^2 )
                   +
                   p_\mu  p_\nu  p_\rho  X_{ 111 }  ( p^2,  M^2,  m_1^2,  m_2^2 )
                   $}
                \Bigr].
    \end{align} 
    $X$ functions can be expressed using Feynman parameter integrals, and  using $B$ functions.
    \begin{align}
        X_{001} ( \scalebox{0.85}{$\displaystyle p^2,  M^2,  m_1^2,  m_2^2 $} )
                                                                         &=    \int_0^1  dx  
                                                                                 \int_0^{1-x}  dy  
                                                                                 \frac{ \frac{ 1 }{ 2 }  x  ( 1  -  x  -  y ) }{ M^2  x  +  m_1^2  y  +  m_2^2  ( 1  -  x  -  y )  -  p^2  x  ( 1  -  x ) }
                                                                                 \nn   \\
                                                                                 \nn   \\
                                                                         &=    \frac{ 1 }{ ( m_1^2  -  m_2^2 )^2 }
                                                                                 \bigl[ B_{001}  ( p^2, m_1^2, M^2 )  -  B_{001}  ( p^2, m_2^2, M^2 ) \bigl]
                                                                                 \nn   \\
                                                                                 &~~~~
                                                                                 -
                                                                                 \frac{ 1 }{ m_1^2  -  m_2^2 }
                                                                                 \left[ \frac{ \del }{ \del  p^2 }  B_{00}  ( p^2, m_2^2, M^2 ) \right],
                                                                                 \\
                                                                                 \nn   \\
                                                                                 \nn   \\
        X_{111} ( \scalebox{0.85}{$\displaystyle p^2,  M^2,  m_1^2,  m_2^2 $} )
                                                                        &=    \int_0^1  dx  \int_0^{1-x}  dy  
                                                                                 \frac{ -  x^3  ( 1  -  x  -  y ) }{ [ M^2  x  +  m_1^2  y  +  m_2^2  ( 1  -  x  -  y )  -  p^2  x  ( 1  -  x ) ]^2 }
                                                                                 \nn   \\
                                                                                 \nn   \\ 
                                                                        &=    \frac{ 1 }{ ( m_1^2  -  m_2^2 )^2 }
                                                                                 \left[ B_{111}  ( p^2, m_1^2, M^2 )  -  B_{111}  ( p^2, m_2^2, M^2 ) \right]
                                                                                 \nn   \\
                                                                                 &~~~~
                                                                                 -
                                                                                 \frac{ 1 }{ m_1^2  -  m_2^2 }
                                                                                 \frac{ \del }{ \del  p^2 }  B_{11}  ( p^2, m_2^2, M^2 ). 
    \end{align}

    \subsection{$Y$ functions}
    \label{sec:loopfun_y}
    The definitions of $Y$ functions ($Y_{1},  Y_{2},  Y_{3}$)  are as follows:
        \begin{align} 
                      \int  \frac{ d^D \ell }{ (2  \pi)^D }  \   \frac{ \ell_\mu }{ [ ( \ell  +  p )^2  -  m_{ \chi }^2 ]  ( \ell^2  -  m_A^2)  [ \ell^2  -  \frac{ m_q^2 }{ x ( 1  -  x ) } ] }
             &=    \frac{ i }{ ( 4  \pi )^2 }  p_\mu  Y_1  ( p^2,  m_\chi^2,  m_A^2,  m_q^2 ),
                      \
                      \\
                      \int  \frac{ d^D \ell }{ (2  \pi)^D }  \  \frac{ \ell_\mu }{ [ ( \ell  +  p )^2  -  m_{ \chi }^2 ]  ( \ell^2  -  m_A^2)  [ \ell^2  -  \frac{ m_q^2 }{ x ( 1  -  x ) } ]^2 }
            &=     \frac{ i }{ ( 4  \pi )^2 }  p_\mu  Y_2  ( p^2,  m_\chi^2,  m_A^2,  m_q^2 ),
                      \
                      \\
                      \int  \frac{ d^D \ell }{ (2  \pi)^D }  \  \frac{ \ell_\mu }{ [ ( \ell  +  p )^2  -  m_{ \chi }^2 ]  ( \ell^2  -  m_A^2)  [ \ell^2  -  \frac{ m_q^2 }{ x ( 1  -  x ) } ]^3 }
            &=     \frac{ i }{ ( 4  \pi )^2 }  p_\mu  Y_3  ( p^2,  m_\chi^2,  m_A^2,  m_q^2 ).
        \end{align}
    $Y$ functions can be expressed using Feynman parameter integrals, and  using $B$, $C$, $D$ functions. 
    \begin{align}
            Y_{1} ( m_\chi^2,  m_\chi^2,  m_A^2,  m_q^2 )    
                                                                        &=    \int_0^1  dy  
                                                                                 \int_0^{1-y}  dz
                                                                                 \frac{ -  2  y }{ [ m_\chi^2  y^2  +  \frac{ m_q^2 }{ x ( 1  -  x ) }  z  +  m_A^2  ( 1  -  y  -  z) ] }                                                                                 
                                                                                 \nn   \\
                                                                                 \nn   \\
                                                                        &=    \frac{ 1 }{ m_A^2  -  \frac{ m_q^2 }{ x  ( 1  -  x ) } }
                                                                                 \left[ B_{1}  ( m_\chi^2,  m_A^2,  m_\chi^2 )  -  B_{1}  ( m_\chi^2,  \scalebox{0.8}{$\displaystyle \frac{ m_q^2 }{ x  ( 1  -  x ) } $},  m_\chi^2 ) \right],
                                                                                 \\                                                                                                                                                                                                                                                                                                                                                                                         
                                                                                 \nn   \\
                                                                                 \nn   \\
        Y_{2} ( m_\chi^2,  m_\chi^2,  m_A^2,  m_q^2 )    
                                                                         &=    \int_0^1  dy  
                                                                                 \int_0^{1-y}  dz
                                                                                 \frac{ 2  y  z }{ [ m_\chi^2  y^2  +  \frac{ m_q^2 }{ x ( 1  -  x ) }  z  +  m_A^2  ( 1  -  y  -  z) ]^2 }
                                                                                 \nn   \\
                                                                                 \nn   \\
                                                                        &=    \frac{ 1 }{ m_A^2  -  \frac{ m_q^2 }{ x  ( 1  -  x ) } }
                                                                                 \left[ Y_{1}  ( m_\chi^2,  m_\chi^2,  m_A^2,  m_q^2 )  -  C_2  ( m_\chi^2,  \scalebox{0.8}{$\displaystyle \frac{ m_q^2 }{ x  ( 1  -  x ) } $},  m_\chi^2 ) \right],
                                                                                 \\                                                                                                                                                                                                                                                                                                                                                                                         
                                                                                 \nn   \\
                                                                                 \nn   \\
        Y_{3} ( m_\chi^2,  m_\chi^2,  m_A^2,  m_q^2 )&
                                                                         \nn   \\
                                                                         =    \int_0^1  dy  
                                                                                 \int_0^{1-y}  dz~&
                                                                                 \frac{ -  4  y  z^2 }{ [ m_\chi^2  y^2  +  \frac{ m_q^2 }{ x ( 1  -  x ) }  z  +  m_A^2  ( 1  -  y  -  z) ]^3 }
                                                                                 \nn   \\
                                                                        =    \frac{ 1 }{ \scalebox{0.8}{$\displaystyle m_A^2 $}  -  \frac{ m_q^2 }{ x  ( 1  -  x ) } }
                                                                                 \Biggl[  Y_{2}&  ( m_\chi^2,  m_\chi^2,  m_A^2,  m_q^2 )
                                                                                         -  D_3 ( \scalebox{0.8}{$\displaystyle 0, 0, m_\chi^2,  m_\chi^2, $}  \scalebox{0.8}{$\displaystyle \frac{ m_q^2 }{ x  ( 1  -  x ) } $},  \scalebox{0.8}{$\displaystyle \frac{ m_q^2 }{ x  ( 1  -  x ) } $},  \scalebox{0.8}{$\displaystyle \frac{ m_q^2 }{ x  ( 1  -  x ) } $},  \scalebox{0.8}{$\displaystyle m_\chi^2 $} ) 
                                                                                 \Biggr].
    \end{align}
    The definitions of $C_2,  D_3$ are in Appendix \ref{sec:loopcd}.

    \subsection{$C_2, D_3$ functions}
    \label{sec:loopcd}
    The definitions of $C_2$ and $D_3$ functions are as follows:
        \begin{align} 
                      \int  \frac{ d^D \ell }{ (2  \pi)^D }  \   \frac{ \ell_\mu }{ ( \ell^2  -  m^2)^2  [ ( \ell  +  p )^2  -  M^2 ] }
             &=    \frac{ i }{ ( 4  \pi )^2 }  p_\mu  C_2  ( p^2,  m^2,  M^2 ),
                      \\
                      \nn   \\
                      \int  \frac{ d^D \ell }{ (2  \pi)^D }  \   \frac{ \ell_\mu }{ ( \ell^2  -  m^2)^3  [ ( \ell  +  p )^2  -  M^2 ] }
            &=     \frac{ i }{ ( 4  \pi )^2 }  p_\mu  D_3  ( p^2,  m^2,  M^2 ).
    \end{align}    
    We have some variations to express $C_2, D_3$ functions~\cite{1501.04161}.
    \begin{align}
        C_2  &  ( p^2,  m^2,  M^2 )    \nn   \\
                                                      &=    \int_0^1  dx  \frac{ x  ( 1  -  x ) }{ m^2  x  +  M^2  x -  p^2  x  ( 1  -  x ) }
                                                        =    \frac{ \del }{ \del  p^2 }  B_0  ( p^2,  m^2,  M^2 ),
                                                              \\
                                                              \nn   \\
        D_3  &  ( p^2,  m^2,  M^2 )    
                                                  \nn   \\
                                                  &=    \int_0^1  dx  \frac{ -  \frac{ 1 }{ 2 }  x^2  ( 1  -  x ) }{ [ m^2  x  +  M^2  x -  p^2  x  ( 1  -  x ) ]^2 }
                                                           \nn   \\
                                                  &=    \frac{ 1 }{ 2 }  \frac{ \del }{ \del  m^2 }  \frac{ \del }{ \del  p^2 }  B_0  ( p^2,  m^2,  M^2 )
                                                           \nn   \\
                                                  &=    \frac{ 1 }{ 2 }  \left( \frac{ \del }{ \del  m^2 } \right)^2  B_0  ( p^2,  m^2,  M^2 )
                                                          \nn   \\
                                                  &=    \frac{ -  p^2  +  m^2  +  M^2 }{ ( p^4  +  m^4  +  M^4  -  2  p^2  m^2  -  2  p^2  M^2  -  2  m^2  M^2 )^2 }
                                                           \nn   \\
                                                           &~~~~~~
                                                           \times
                                                           \left[
                                                               \scalebox{0.9}{$\displaystyle
                                                               ( M^2  -  m^2  +  p^2 )  \bigl[ -B_0  ( p^2,  m^2,  M^2 )  +  B_0  ( 0,  m^2,  m^2 )  +  2 \bigr]  -  2  M^2  \log  \frac{M^2}{m^2}
                                                               $}
                                                           \right]
                                                           \nn   \\
                                                           &~~
                                                           +  \frac{ 1 }{ 2 }
                                                           \frac{ 1 }{ ( p^4  +  m^4  +  M^4  -  2  p^2  m^2  -  2  p^2  M^2  -  2  m^2  M^2 ) }
                                                           \nn   \\
                                                           &~~~~~~
                                                           \times
                                                           \left[
                                                               \scalebox{0.9}{$\displaystyle
                                                               -B_0  ( p^2,  m^2,  M^2 )  +  B_0  ( 0,  m^2,  m^2 )  +  2  -  ( p^2  -  m^2  -  M^2 )  \frac{ \del }{ \del  p^2 }  B_0  ( p^2,  m^2,  M^2 )
                                                               $}
                                                           \right].
    \end{align}

    \subsection{$\del  F( m_a^2 )  /  \del  m_a^2$ in loop functions}
    \label{sec:dfdm2}
    $ \del  Y_n  ( p^2,  m_\chi^2,  m_a^2,  m_q^2 )  /  \del  m_a^2  \  ( n  =  1,  2,  3 )$ can be expressed in the loop functions introduced above. 
    \begin{align}
        \frac{ \del }{ \del  m_a^2 }  Y_1  ( \scalebox{0.85}{$\displaystyle p^2,  m_\chi^2,  m_a^2,  m_q^2 $} ) 
                =&    
                        \frac{ 1 }{ m_a^2  -  \frac{ m_q^2 }{ x  ( 1  -  x ) } }
                        \left[ 
                              \frac{ \del }{ \del  m_\chi^2 }  B_0  ( \scalebox{0.85}{$\displaystyle m_\chi^2,  m_a^2,  m_\chi^2 $} )  -  Y_1  ( \scalebox{0.85}{$\displaystyle p^2,  m_\chi^2,  m_a^2,  m_q^2 $} ) 
                        \right],
                        \\
                        \nn   \\
        \frac{ \del }{ \del  m_a^2 }  Y_2  ( \scalebox{0.85}{$\displaystyle p^2,  m_\chi^2,  m_a^2,  m_q^2 $} ) 
                =&
                        \frac{ 1 }{ m_a^2  -  \frac{ m_q^2 }{ x  ( 1  -  x ) } }
                        \left[ 
                            \frac{ \del }{ \del  m_a^2 }  Y_1  ( \scalebox{0.85}{$\displaystyle  p^2,  m_\chi^2,  m_a^2,  m_q^2 $} )  -  Y_2  ( \scalebox{0.85}{$\displaystyle p^2,  m_\chi^2,  m_a^2,  m_q^2 $} )
                        \right],
                        \\
                        \nn   \\
        \frac{ \del }{ \del  m_a^2 }  Y_3  ( \scalebox{0.85}{$\displaystyle p^2,  m_\chi^2,  m_a^2,  m_q^2 $} ) 
                =&
                        \frac{ 1 }{ m_a^2  -  \frac{ m_q^2 }{ x  ( 1  -  x ) } }
                        \left[ 
                            \frac{ \del }{ \del  m_a^2 }  Y_2  ( \scalebox{0.85}{$\displaystyle p^2,  m_\chi^2,  m_a^2,  m_q^2 $} )  -  Y_3  ( \scalebox{0.85}{$\displaystyle p^2,  m_\chi^2,  m_a^2,  m_q^2 $} )
                        \right].
    \end{align}
    The expression for $ \del  F( m_a^2 )  /  \del  m_a^2 $ using the loop functions is 
    \begin{align}
        \frac{ \del  F  (m_a^2 ) }{ \del  m_a^2 }
                                      =    \int_0^1  dx
                                            \biggl\{
                                                          &
                                                           3  \frac{ \del }{ \del  m_a^2 }  Y_1  ( p^2,  m_\chi^2,  m_a^2,  m_q^2 ) 
                                                           \nn   \\
                                                          &  -  m_q^2  \frac{ ( 2  +  5  x  -  5  x^2 ) }{ x^2 ( 1  -  x )^2 }  \frac{ \del }{ \del  m_a^2 }  Y_2  ( p^2,  m_\chi^2,  m_a^2,  m_q^2 )
                                                          \nn   \\
                                                          &  -  2  m_q^4  \frac{ ( 1  -  2  x  +  2  x^2 ) }{ x^3 ( 1  -  x )^3 }  \frac{ \del }{ \del  m_a^2 }  Y_3  ( p^2,  m_\chi^2,  m_a^2,  m_q^2 )
                                            \biggr\}.
    \end{align}


\begin{thebibliography}{99}

\bibitem{1705.03380} 
  D.~S.~Akerib {\it et al.} [LUX Collaboration],
  Phys.\ Rev.\ Lett.\  {\bf 118}, no. 25, 251302 (2017)
  doi:10.1103/PhysRevLett.118.251302
  [arXiv:1705.03380 [astro-ph.CO]].


\bibitem{Cui:2017nnn} 
  X.~Cui {\it et al.} [PandaX-II Collaboration],
  Phys.\ Rev.\ Lett.\  {\bf 119}, no. 18, 181302 (2017)
  doi:10.1103/PhysRevLett.119.181302
  [arXiv:1708.06917 [astro-ph.CO]].


\bibitem{1805.12562} 
  E.~Aprile {\it et al.} [XENON Collaboration],
  Phys.\ Rev.\ Lett.\  {\bf 121}, no. 11, 111302 (2018)
  doi:10.1103/PhysRevLett.121.111302
  [arXiv:1805.12562 [astro-ph.CO]].


\bibitem{1609.09079} 
  M.~Escudero, A.~Berlin, D.~Hooper and M.~X.~Lin,
  JCAP {\bf 1612}, 029 (2016)
  doi:10.1088/1475-7516/2016/12/029
  [arXiv:1609.09079 [hep-ph]].


\bibitem{1612.06462} 
  M.~Escudero, D.~Hooper and S.~J.~Witte,
  JCAP {\bf 1702}, no. 02, 038 (2017)
  doi:10.1088/1475-7516/2017/02/038
  [arXiv:1612.06462 [hep-ph]].


\bibitem{1404.3716} 
  S.~Ipek, D.~McKeen and A.~E.~Nelson,
  Phys.\ Rev.\ D {\bf 90}, no. 5, 055021 (2014)
  doi:10.1103/PhysRevD.90.055021
  [arXiv:1404.3716 [hep-ph]].


\bibitem{1509.01110} 
  J.~M.~No,
  Phys.\ Rev.\ D {\bf 93}, no. 3, 031701 (2016)
  doi:10.1103/PhysRevD.93.031701
  [arXiv:1509.01110 [hep-ph]].


\bibitem{1611.04593} 
  D.~Goncalves, P.~A.~N.~Machado and J.~M.~No,
  Phys.\ Rev.\ D {\bf 95}, no. 5, 055027 (2017)
  doi:10.1103/PhysRevD.95.055027
  [arXiv:1611.04593 [hep-ph]].


\bibitem{1701.07427} 
  M.~Bauer, U.~Haisch and F.~Kahlhoefer,
  JHEP {\bf 1705}, 138 (2017)
  doi:10.1007/JHEP05(2017)138
  [arXiv:1701.07427 [hep-ph]].


\bibitem{1705.09670} 
  P.~Tunney, J.~M.~No and M.~Fairbairn,
  Phys.\ Rev.\ D {\bf 96}, no. 9, 095020 (2017)
  doi:10.1103/PhysRevD.96.095020
  [arXiv:1705.09670 [hep-ph]].


\bibitem{1711.02110} 
  G.~Arcadi, M.~Lindner, F.~S.~Queiroz, W.~Rodejohann and S.~Vogl,
  JCAP {\bf 1803}, no. 03, 042 (2018)
  doi:10.1088/1475-7516/2018/03/042
  [arXiv:1711.02110 [hep-ph]].


\bibitem{1712.03874} 
  P.~Pani and G.~Polesello,
  Phys.\ Dark Univ.\  {\bf 21}, 8 (2018)
  doi:10.1016/j.dark.2018.04.006
  [arXiv:1712.03874 [hep-ph]].


\bibitem{1803.01574} 
  N.~F.~Bell, G.~Busoni and I.~W.~Sanderson,
  JCAP {\bf 1808}, no. 08, 017 (2018)
  doi:10.1088/1475-7516/2018/08/017
  [arXiv:1803.01574 [hep-ph]].


\bibitem{1804.02120} 
  T.~Li,
  Phys.\ Lett.\ B {\bf 782}, 497 (2018)
  doi:10.1016/j.physletb.2018.05.073
  [arXiv:1804.02120 [hep-ph]].


\bibitem{1408.4929} 
  K.~Ghorbani,
  JCAP {\bf 1501}, 015 (2015)
  doi:10.1088/1475-7516/2015/01/015
  [arXiv:1408.4929 [hep-ph]].


\bibitem{1411.1335} 
  T.~Abe, R.~Kitano and R.~Sato,
  Phys.\ Rev.\ D {\bf 91}, no. 9, 095004 (2015)
  Erratum: [Phys.\ Rev.\ D {\bf 96}, no. 1, 019902 (2017)]
  doi:10.1103/PhysRevD.96.019902, 10.1103/PhysRevD.91.095004
  [arXiv:1411.1335 [hep-ph]].


\bibitem{1701.04131} 
  S.~Baek, P.~Ko and J.~Li,
  Phys.\ Rev.\ D {\bf 95}, no. 7, 075011 (2017)
  doi:10.1103/PhysRevD.95.075011
  [arXiv:1701.04131 [hep-ph]].


\bibitem{1702.07236} 
  T.~Abe,
  Phys.\ Lett.\ B {\bf 771}, 125 (2017)
  doi:10.1016/j.physletb.2017.05.048
  [arXiv:1702.07236 [hep-ph]].


\bibitem{1307.5458} 
  J.~Billard, L.~Strigari and E.~Figueroa-Feliciano,
  Phys.\ Rev.\ D {\bf 89}, no. 2, 023524 (2014)
  doi:10.1103/PhysRevD.89.023524
  [arXiv:1307.5458 [hep-ph]].


\bibitem{1512.07501} 
  E.~Aprile {\it et al.} [XENON Collaboration],
  JCAP {\bf 1604}, no. 04, 027 (2016)
  doi:10.1088/1475-7516/2016/04/027
  [arXiv:1512.07501 [physics.ins-det]].


\bibitem{1611.05525} 
  M.~Szydagis [LUX and LZ Collaborations],
  PoS ICHEP {\bf 2016}, 220 (2016)
  doi:10.22323/1.282.0220
  [arXiv:1611.05525 [astro-ph.CO]].


\bibitem{1606.07001} 
  J.~Aalbers {\it et al.} [DARWIN Collaboration],
  JCAP {\bf 1611}, 017 (2016)
  doi:10.1088/1475-7516/2016/11/017
  [arXiv:1606.07001 [astro-ph.IM]].


\bibitem{Shifman:1978zn} 
  M.~A.~Shifman, A.~I.~Vainshtein and V.~I.~Zakharov,
  Phys.\ Lett.\  {\bf 78B}, 443 (1978) 
  doi:10.1016/0370-2693(78)90481-1. 


\bibitem{PhysRev.D41.3421} 
  V.~D.~Barger, J.~L.~Hewett and R.~J.~N.~Phillips,
  Phys.\ Rev.\ D {\bf 41}, 3421 (1990) 
  doi:10.1103/PhysRevD.41.3421. 


\bibitem{hep-ph/9401311} 
  Y.~Grossman,
  Nucl.\ Phys.\ B {\bf 426}, 355 (1994)
  doi:10.1016/0550-3213(94)90316-6
  [hep-ph/9401311].


\bibitem{0902.4665} 
  M.~Aoki, S.~Kanemura, K.~Tsumura and K.~Yagyu,
  Phys.\ Rev.\ D {\bf 80}, 015017 (2009)
  doi:10.1103/PhysRevD.80.015017
  [arXiv:0902.4665 [hep-ph]].


\bibitem{PhysRev.D15.1958} 
  S.~L.~Glashow and S.~Weinberg,
  Phys.\ Rev.\ D {\bf 15}, 1958 (1977) 
  doi:10.1103/PhysRevD.15.1958. 


\bibitem{1508.05716} 
  X.~Liu, L.~Bian, X.~Q.~Li and J.~Shu,
  Nucl.\ Phys.\ B {\bf 909}, 507 (2016)
  doi:10.1016/j.nuclphysb.2016.05.027
  [arXiv:1508.05716 [hep-ph]].


\bibitem{ATLAS:2018doi} 
  The ATLAS collaboration [ATLAS Collaboration],
  ATLAS-CONF-2018-031.


\bibitem{1809.10733} 
  A.~M.~Sirunyan {\it et al.} [CMS Collaboration],
  [arXiv:1809.10733 [hep-ex]].


\bibitem{1702.04571} 
  M.~Misiak and M.~Steinhauser,
  Eur.\ Phys.\ J.\ C {\bf 77}, no. 3, 201 (2017)
  doi:10.1140/epjc/s10052-017-4776-y
  [arXiv:1702.04571 [hep-ph]].


\bibitem{1712.06518} 
  M.~Aaboud {\it et al.} [ATLAS Collaboration],
  JHEP {\bf 1803}, 174 (2018)
  Erratum: [JHEP {\bf 1811}, 051 (2018)]
  doi:10.1007/JHEP11(2018)051, 10.1007/JHEP03(2018)174
  [arXiv:1712.06518 [hep-ex]].


\bibitem{CMS-PAS-HIG-18-005} 
  CMS Collaboration [CMS Collaboration],
  CMS-PAS-HIG-18-005.


\bibitem{CMS-PAS-HIG-17-031} 
  CMS Collaboration [CMS Collaboration],
  CMS-PAS-HIG-17-031.


\bibitem{Hisano:2017jmz} 
  J.~Hisano,
  arXiv:1712.02947 [hep-ph].


\bibitem{1305.0237} 
  G.~Belanger, F.~Boudjema, A.~Pukhov and A.~Semenov,
  Comput.\ Phys.\ Commun.\  {\bf 185}, 960 (2014)
  doi:10.1016/j.cpc.2013.10.016
  [arXiv:1305.0237 [hep-ph]].


\bibitem{Pumplin:2002vw} 
  J.~Pumplin, D.~R.~Stump, J.~Huston, H.~L.~Lai, P.~M.~Nadolsky and W.~K.~Tung,
  JHEP {\bf 0207}, 012 (2002)
  doi:10.1088/1126-6708/2002/07/012
  [hep-ph/0201195].


\bibitem{Novikov:1983gd} 
  V.~A.~Novikov, M.~A.~Shifman, A.~I.~Vainshtein and V.~I.~Zakharov,
  Fortsch.\ Phys.\  {\bf 32}, 585 (1984).


\bibitem{1007.2601} 
  J.~Hisano, K.~Ishiwata and N.~Nagata,
  Phys.\ Rev.\ D {\bf 82}, 115007 (2010)
  doi:10.1103/PhysRevD.82.115007
  [arXiv:1007.2601 [hep-ph]].


\bibitem{1501.04161} 
  T.~Abe and R.~Sato,
  JHEP {\bf 1503}, 109 (2015)
  doi:10.1007/JHEP03(2015)109
  [arXiv:1501.04161 [hep-ph]].


\bibitem{1502.01589} 
  P.~A.~R.~Ade {\it et al.} [Planck Collaboration],
  Astron.\ Astrophys.\  {\bf 594}, A13 (2016)
  doi:10.1051/0004-6361/201525830
  [arXiv:1502.01589 [astro-ph.CO]].


\bibitem{Hahn:1998yk} 
  T.~Hahn and M.~Perez-Victoria,
  Comput.\ Phys.\ Commun.\  {\bf 118}, 153 (1999)
  doi:10.1016/S0010-4655(98)00173-8
  [hep-ph/9807565].


\bibitem{Hisano:2004pv} 
  J.~Hisano, S.~Matsumoto, M.~M.~Nojiri and O.~Saito,
  Phys.\ Rev.\ D {\bf 71}, 015007 (2005)
  doi:10.1103/PhysRevD.71.015007
  [hep-ph/0407168].


\bibitem{1004.4090} 
  J.~Hisano, K.~Ishiwata and N.~Nagata,
  Phys.\ Lett.\ B {\bf 690}, 311 (2010)
  doi:10.1016/j.physletb.2010.05.047
  [arXiv:1004.4090 [hep-ph]].


\bibitem{1504.00915} 
  J.~Hisano, K.~Ishiwata and N.~Nagata,
  JHEP {\bf 1506}, 097 (2015)
  doi:10.1007/JHEP06(2015)097
  [arXiv:1504.00915 [hep-ph]].

\end{thebibliography}
\end{document}